\documentclass[a4paper, 12pt]{article}

\usepackage{epsfig,amssymb,euscript,xspace, color}
\usepackage{amsmath,jheppub_0,mathtools}
\def\bea{\begin{eqnarray}}
\def\eea{\end{eqnarray}}
\def\be{\begin{equation}}
\def\ee{\end{equation}}



\def\cL{{\cal L}}

\def\cL{{\cal L}}

\begin{document}

\title{Rotating Black Holes in $AdS$, Extremality and Chaos}
\author[a]{Avik Banerjee}
\author[b]{ Arnab Kundu}
\author[b]{Rohan R. Poojary}
\affiliation[a]{Harish-Chandra Research Institute, \\
Homi Bhaba National Institute (HBNI),\\
Chhatnag Road, Jhusi, Allahabad 211019, India. }
\affiliation[b]{Theory Division, Saha Institute of Nuclear Physics, \\
Homi Bhaba National Institute (HBNI), \\
1/AF, Bidhannagar, Kolkata 700064, India. }
\emailAdd{avikbanerjee@hri.res.in,~arnab.kundu@saha.ac.in,~rohan.poojary@saha.ac.in}

\abstract{Extremal black holes have vanishing Hawking temperatures. In this article, we argue that for asymptotically AdS black holes, at extremality, a particular class of correlators in the dual CFT can exhibit exponential, maximally chaotic growth with a non-vanishing temperature. Our approach, at extremality, is two-fold. First, we geometrically investigate the modes that are responsible for chaos. Secondly, we study the dynamics of a probe string to capture chaos in worldsheet correlators.
For rotating BTZ at extremality, the corresponding Lyapunov exponent is determined by the left-moving temperature. In higher dimensional AdS-Kerr geometries, on the other hand, the corresponding Lyapunov exponent becomes a non-trivial function of the Frolov-Thorne temperatures.}

\maketitle

\section{Introduction}
The study of black holes $via$ holography has allowed us uncover interesting features of quantum gravity, especially in $AdS$ space-times. It allows the study of large-$N$ QFTs to be related to the (thermo)dynamics of black holes in $AdS$.   
It was discovered that certain thermal large-$N$ unitary systems can scramble the information of an initial localized perturbation amongst its microscopic $d.o.f.$ exponentially faster than others \cite{Hayden:2007cs}. These were termed as fast scramblers and they possessed  a scrambling time $t*\sim \log N$. Further, black holes where conjectured to be fastest amongst the fast scramblers \cite{Sekino:2008he}.
\\\\
In the thermal system, scrambling time as measured by how fast $C(t)=\left\langle[V(t),W(0)]\right\rangle_\beta^2$ grows in early time is a good diagnostic of its chaotic behaviour. For fast scramblers it was seen that $C(t)\sim e^{\lambda_L}t$, where the Lyapunov exponent $\lambda_L$ measures the growth in OTOC $\left\langle V(t)W(0)V(t)W(0)\right\rangle_\beta$ occurring in $\left\langle[V(t),W(0)]\right\rangle_\beta^2$.
The holographic computation of $\lambda_L$ for Schwarzchild-$AdS$ black holes revealed $\lambda_L=2\pi/\beta$ \cite{Shenker:2013pqa,Shenker:2013yza,Shenker:2014cwa} with stringy corrections only decreasing it. 
Validating the fast scrambling conjecture it was shown that generic thermal large-$N$ systems satisfy a bound on chaos with $\lambda_L\le 2\pi/\beta$ \cite{Maldacena:2015waa}. 
\\\\
Similar chaotic behaviour was obtained for a large-$N$ system dual to a string probing a Schwarzchild-$AdS_3$ stretched from the boundary to the horizon \cite{deBoer:2017xdk}. The string is governed by the Nambu-Goto action and the end point of the string is dual to a quark in a thermal large-$N$ system \cite{Dubovsky:2017cnj,Dubovsky:2012wk}.
Here although the string world-sheet sees the ambient horizon it is entirely the string dynamics which gives rise to the chaotic behaviour. 
\\\\
The analysis of the now famous SYK model \cite{Sachdev:2010um,Polchinski:2016xgd, Kitaev:2014talk,Maldacena:2016hyu} revealed a similar behaviour close to small temperatures, although being a solvable model in large-$N$. Thus suggesting that it may be a good model to simulate the chaotic dynamics of black holes in gravity. Many properties of these models have since been under extensive investigation\cite{Kitaev:2017awl,Sonner:2017hxc,Eberlein:2017wah,Gross:2017vhb,Stanford:2017thb}. The low energy dynamics of SYK-like models is governed by the Schwarzian action for the time re-parametrizations which is a symmetry of its  zero temperature fixed point. This suggested that atleast close to extremality a similar behaviour may hold for $AdS$-black holes. The Jackiw-Teitelboim (JT) model studied in \cite{Jensen:2016pah,Maldacena:2016upp} essentially captures this behaviour for small temperatures close to extremality. This model is a 2 dimensional gravity description for deviations away from extremality where black holes are known to have an $AdS_2$ factor in their near horizon (throat) limit \cite{Kunduri:2007vf}. 
 Similar Schwarzian dynamics has also been uncovered for explaining the late time behaviour of probe strings in Schwarzchild-$AdS_3$ \cite{Banerjee:2018kwy,Banerjee:2018twd}.    
\\\\ However, study of rotating geometries with regards to computing their Lyapunov exponents reveals some interesting unexplored aspects.  
It was shown for generic rotating BTZ metrics that the Lyapunov exponent can be $>2\pi/\beta$. Here apparently 2 Lyapunov indices were found $\lambda_L^+<2\pi/\beta<\lambda_L^-$ \cite{Poojary:2018eszz,Jahnke:2019gxr}, where $\lambda_L^\pm=r_+\mp r_-$ with $r_-(r_+)$ being the inner(outer) horizons. This suggests that in presence of a chemical potential $\beta\mu$ the bound derived on thermal large-$N$ QFTs may be modified. This was indeed shown to be true for certain states in \cite{Halder:2019ric} by modifying the arguments of \cite{Maldacena:2016hyu}  due to the presence of a chemical potential for a continuous global $U(1)$ symmetry. The study of translational symmetry on the chaos bound was studied in \cite{Mezei:2019dfv}. The above recent observations taken in tandem with the fast scrambling conjecture seems to suggest that rotating black holes in higher dimensions may also exhibit a similar behaviour. It is interesting to note that the one dimensional solvable SYK-like models do not seem to saturate such a modified bound in presence of a chemical potential \cite{Bhattacharya:2017vaz,Yoon:2017nig}. Higher dimensional versions of the SYK-like models have also been under investigations \cite{Berkooz:2016cvq}.
\\\\
An interesting consequence due to the presence of chemical potential is extremal chaos, as can be seen in the case of BTZ in \cite{Poojary:2018eszz,Jahnke:2019gxr}. The near horizon dynamics as captured by the JT model does not seem to account for this effect. However, it correctly reproduces the thermodynamics of charged and rotating black holes close to extremality which has been thoroughly verified in \cite{Nayak:2018qej,Moitra:2019bub,Moitra:2018jqs}. For past works on dimensional reduction refer to \cite{Gouteraux:2011qh,Castro:2008ms}. It would be therefore interesting to see how the JT model accounts for the chaos causing modes in the near horizon region with regards to Lyapunov index as seen at the conformal boundary of rotating BTZ. This would also be useful in higher dimensions as holographic analysis of computing the Lyapunov exponent using known techniques in literature would be a cumbersome task in rotating black hole geometries in dimensions $>3$. This paper attempts to take the first steps in this direction. Also recently there have been studies in this regard \cite{Ghosh:2019rcj}.
\\\\

We organize the paper as follows:
In section 2 we review briefly the various computations used to deduce $\lambda_L$ holographically and the change in $\lambda_L$ in generic large-$N$ thermal systems with chemical potential.
In section 3 we analyse the near horizon region of BTZ close to extremality and indeed find that the thermal modes captured by the JT model contribute a Lyapunov index $\lambda_L=2\pi T_H$\footnote{$T_H=\beta^{-1}$.}. At extremality the thermal modes are essentially similar to the right moving modes seen at the boundary of BTZ which see a zero temperature and therefore contribute $\lambda_L^+=0$. 
In section 4 we see that the left moving extremal modes at extremality in the near horizon region are seen to give rise to a $\lambda^-_L=2r_+$ in tandem with the analysis at conformal boundary of BTZ. However we see while deriving the near horizon effective action that the presence of the thermal modes is necessary in deducing this effect. 
In section 5 we supplement the BTZ analysis by studying the temperature dependence of a probe string in the BTZ geometry on the angular velocity of its end point.  
In Section 6
we then analyse the near horizon region of extremal Kerr-$AdS_4$ and find the thermal modes which are captured by the JT analysis. Drawing parlance with the BTZ near horizon analysis we look for similar  diffeomorphisms of the near horizon Kerr metric that can give rise to extremal chaos. We find that these are precisely the `large' diffeomorphisms studied in the Kerr/CFT correspondence\footnote{We do not need the Kerr/CFT correspondence to  motivate extremal chaos.}. We also see that one can easily find the temperature as seen by the extremal modes  by comparing the warped $AdS_3$ in the throat region of extremal Kerr-$AdS_4$ to the near horizon extremal BTZ metric. 
We similarly study the temperature as seen by the probe string in extreme Kerr-$AdS_4$ as function of its end points angular velocity at the conformal boundary.
\\\\
We note some important list of references before beginning below:
Various aspects of the SYK-like models and the JT gravity have been studied to a great detail in the recent years \cite{Polchinski:2016xgd,Jevicki:2016bwu,Jevicki:2016ito,Engelsoy:2016xyb,Bagrets:2016cdf,Cvetic:2016eiv,Gu:2016oyy,Berkooz:2016cvq,Garcia-Garcia:2016mno,Fu:2016vas,Witten:2016iux,Gurau:2016lzk,Klebanov:2016xxf,Choudhury:2017tax,Davison:2016ngz,Peng:2016mxj,Krishnan:2016bvg,Krishnan:2017lra,Krishnan:2017txw,Turiaci:2017zwd,Li:2017hdt,Gurau:2017xhf,Mandal:2017thl,Bonzom:2017pqs,Gross:2017hcz,Stanford:2017thb,Maldacena:2017axo,Das:2017hrt,Das:2017pif,Das:2017wae,Mertens:2017mtv,Gross:2017vhb,Taylor:2017dly,Giombi:2018qgp,Giombi:2017dtl,Sonner:2017hxc,Bulycheva:2017ilt,Grumiller:2017qao,Gross:2017aos,Narayan:2017qtw,Gonzalez:2018enk,Roberts:2018mnp,Benedetti:2018goh,Dhar:2018pii,Gaikwad:2018dfc,Klebanov:2018nfp,Gharibyan:2018jrp,Maldacena:2018lmt,Harlow:2018tqv,Brown:2018bms,Lam:2018pvp,Bena:2018bbd,Saad:2018bqo,Gubser:2018yec,Larsen:2018iou,Blommaert:2019hjr,Chang:2018sve,Gur-Ari:2018okm,Goel:2018ubv,Lin:2018xkj,Castro:2018ffi,Liu:2018jhs,Pakrouski:2018jcc,Blake:2018leo,Yang:2018gdb,Brown:2018kvn,Kolekar:2018sba,Larsen:2018cts,Bhattacharya:2018fkq,Murugan:2018fdj,Goto:2018iay,Kim:2019upg,Sachdev:2019bjn,Nayak:2019khe,Sun:2019mms,Saad:2019lba,Mertens:2019bvy,Maldacena:2019cbz,Guo:2019csw,Mertens:2019tcm,Susskind:2019ddc,Lin:2019qwu,Iliesiu:2019xuh,Klebanov:2019jup,Sun:2019yqp,Choi:2019bmd}. An extensive list of references for works on SYK-like models and a through review of the same can be found in \cite{Rosenhaus:2019mfr}. Aspects of JT gravity have also been reviewed in \cite{Sarosi:2017ykf}.
Rotating  Kerr geometries have been studied in detail with regards to their near horizon properties at extremality. For an account of interesting aspects of Kerr geometries and Kerr/CFT correspondence please refer to  \cite{Guica:2010ej,Guica:2008mu,Hartman:2008pb,Hartman:2009nz,Bredberg:2009pv,Castro:2010fd,Porfyriadis:2014fja,Hadar:2014dpa,Hadar:2015xpa,Porfyriadis:2016gwb,Compere:2012jk,Sen:2005wa,Astefanesei:2006dd,Goldstein:2005hq,Bardeen:1999px,Hawking:1998kw,Gibbons:2004ai,Gibbons:2004js,Hawking:1999dp,Caldarelli:1999xj,Lemos:1996bq}. Recently there have also been interesting investigations of $nAdS_2/CFT_1$ and large diffeomorphisms in the near horizon region in extreme Kerr \cite{Castro:2019crn}. The case of the BTZ with gravitational Chern-Simons term and it's effect in the near horizon region was also recently studied using $nAdS_2/CFT_1$ \cite{Castro:2019vog}.   The case of rotating black holes dual to holographic CFTs and strong cosmic censorship was recently studied in \cite{Balasubramanian:2019qwk}.

\section{Chaos in BTZ}
The rotating BTZ solution is generically written as 
\bea
\frac{ds^2}{l^2}&=&\frac{\rho^2d\rho^2}{(\rho^2-r_+^2)(\rho^2-r_-^2)}-\frac{(\rho^2-r_+^2)(\rho^2-r_-^2)dt^2}{\rho^2}+r^2\left(d\phi-\frac{r_+r_-}{\rho^2}dt\right)^2\cr&&\cr
&&\hspace{-1.2cm}{\rm having}\hspace{0.5cm}M=r_+^2+r_-^2,\,\,J=2lr_+r_-.
\label{BTZ_metric_wiki}
\eea
with $r_\pm$ labelling in outer and inner horizons respectively. It can also be cast in the  Fefferman Graham gauge to be precisely  
\bea
\frac{ds^2}{l^2}&=&\frac{dr^2}{r^2}-\frac{r^2dx^+dx^-}{4}+\frac{1}{4}\left(T_{++}dx^{+2}+T_{--}dx^{-2}\right)-\frac{1}{4r^2}T_{++}T_{--}dx^+dx^-,
\label{Banados_metric}
\eea 
where $T_{\pm\pm}=(r_+\mp r_-)^2$, here the location of the horizon can be computed from computing where the area of constant $r$ hyper-surface vanishes $i.e.$ $r_h=\sqrt{r_+^2-r_-^2}$. 
\\\\\
The Fefferman-Graham coordinates are especially useful since one can readily write down the most generic solution to Einstein's equations which are locally $AdS_3$ by simply promoting $T_\pm\rightarrow T_\pm(x^\pm)$\footnote{Here we assume Dirichlet boundary conditions have been imposed}. As there are no bulk degrees of freedom for gravity in 3-dim all these solutions can be reached from an arbitrary metric of the form \eqref{Banados_metric} $via$ local coordinate transformations $i.e.$ finite diffeomorphisms; see for $e.g.$\cite{Roberts:2012aq}\& Appendix \ref{Appendix_A}.   The infinitesimal forms of these diffeomorphisms form the well known Brown-Henneaux asymptotic symmetries of $AdS_3$. We call them Penrose-Brown-Henneaux(PBH) diffeomorphisms here.  These are the $Vir\otimes Vir$  with central charge $c=3l/2G_N$ where $l$ is the $AdS_3$ length scale as seen in \eqref{BTZ_metric_wiki} and \eqref{Banados_metric}.
\\\\
For stationary BTZ it was shown in \cite{Poojary:2018eszz} that the 4pt. function of 2 scalar primaries $V,W$ dual to minimally coupled scalar fields $\phi_V,\phi_W$ of the form $\left\langle  VWVW\right\rangle$ receives corrections from the bulk on-shell action arising from \eqref{Banados_metric} with $T_{\pm\pm}\rightarrow T_{\pm\pm}(x^\pm)$
\bea
&&S^{on-shell}_{(3)}=\frac{l}{32\pi G_N} \int_\partial \sqrt{T_{++}T_{--}}.\cr&&\cr
&&T_{\pm,\pm}=-2\,{\rm Sch}[X^\pm,x^\pm]+\left(\frac{2\pi}{\beta_\pm}\right)^2X'^{\pm2}
\label{onshell_BTZ_action}
\eea
Here the 3-dim family of solutions are parametrized by $x^\pm\rightarrow X^\pm(x^\pm)$ conformal transformations on the boundary of BTZ.
For $OTOC$s of the form $\langle V(t,0)W(0,\phi)V(t,0)W(0,\phi)\rangle$ we find that there are 2 Lyapunov exponents, $\lambda_L^\pm=\frac{2\pi}{\beta_\pm}=r_+\mp r_-$ of which one is not only greater that $2\pi T_H=\frac{r_+^2-r_-^2}{r^+}$ but also seems to survive when we take the extremal limit $i.e.$ $\lambda_L^-=r_++r_-\overset{\rm ext}{\longrightarrow}2r_+$. This method was also carried out in the first order formalism of $AdS_3$ gravity as a difference of 2 Chern-Simons theories in $sl(2,\mathbb{R})$ in \cite{Jahnke:2019gxr}.
\\\\
The extremal limit is reached when the inner and outer horizons coincide $i.e.$ $r_+-r_-=2\epsilon\rightarrow 0$. Expanding $2\pi T_\pm=\lambda^\pm_L$ and $2\pi T_H$ in terms of $\epsilon$ we find
\be
2\pi T_H=\frac{r_+^2-r_-^2}{r_+}=4\epsilon -\frac{4\epsilon^2}{r_+},\hspace{0.5cm}\lambda_L^-=2(r_+-\epsilon),\hspace{0.5cm}\lambda_L^+=2\epsilon.
\ee
The corrections to probe computation of $\langle VWVW \rangle$ could also be carried out by taking into account shock-wave back reactions for late times of the scale $t\gg\beta$ as was first done in \cite{Shenker:2014cwa} for non-rotating BTZ. Here one utilises the Thermo-Field Double (TFD) description of the BTZ black hole. The 4pt. function is computed by considering Eikonal scattering of scalar fields $\phi_V,\phi_W$. Since one is interested in large time dynamics this prescription translates to considering scattering of scalar fields in a backreacted geometry very close to the horizon. This is consistent with the fact that since one is interested in computing late time effects of small perturbations to the bulk, the dominant effect from gravitational back reactions  would arise from the region closest to the horizon. Indeed, the scrambling time of the bulk theory is the smallest time for small perturbations to the bulk to have a measurable effect as seen from the boundary theory.
\\\\
The `in' and `out' states for the Eikonal scattering are constructed on the horizons labelled by (Kruskal) bulk co-ordinates $u=0$ and $v=0$ as follows:
\bea
&&|{\rm in}\rangle =V(t_3)W(t_4)|{\rm TFD}\rangle_{\rm in}=\int d\phi'_3 d\phi'_4 dp^u_3dp^v_4\,\,\psi_3(p_3^u,\phi'_3)\psi_4(p_4^v,\phi'_4)|p_3^u,\phi'_3\rangle\otimes|p_4^v,\phi'_4\rangle\cr
&&\hspace{-0.2cm}|{\rm out}\rangle =V(t_1)^\dagger W(t_2)^\dagger|{\rm TFD}\rangle_{\rm out}=\int d\phi'_1 d\phi'_2 dp^u_1dp^v_2\,\,\psi_1^\dagger(p_1^u,\phi'_1)\psi_2^\dagger(p_2^v,\phi'_2)|p_1^u,\phi'_1\rangle\otimes|p_2^v,\phi'_2\rangle.\cr&&
\eea
Here the $\psi$s for any bulk  operator $\mathcal{O}$ are the bulk to boundary propagators computed in the corresponding  BTZ geometry at the required horizon as:
\be
\psi(p^u,\phi)=\int dv e^{-ip_v v}\langle \phi_\mathcal{O}(u,v,\phi)\mathcal{O}(t)\rangle|_{u=0}.
\ee
The Eikonal scattering for these states then results in the `in' and `out' states to differ by a phase factor for elastic scattering
\be
(|p_1^u,\phi'_1\rangle\otimes|p_2^v,\phi'_2)_{\rm out}\thickapprox e^{i\delta(s,\phi'_1-\phi'_2)}(|p_3^u,\phi'_1\rangle\otimes|p_4^v,\phi'_2\rangle)_{\rm in}+ |{\rm inelastic}\rangle.
\ee
The final answer is the obtained by evaluating
\be
\langle{\rm out}|{\rm in}\rangle=\,\,_{\rm out}\langle TFD|V(t_1)W(t_2)V(t_3)W(t_4)|{\rm TFD}\rangle_{\rm in}
\ee    
For rotating BTZ it was shown in \cite{Jahnke:2019gxr} that such a computation indeed gives rise to 2 Lyapunov exponents $\lambda_\pm=r_+\mp r_-$. 
This method although gives the full form of the answer for late times doesn't reveal the mechanism behind the dynamics which results in maximal chaos. It is also computationally involved as one needs to compute bulk-to-boundary propagators for rotating geometries which resist a simple coordinate system in their Kruskal extensions. Bulk to boundary propagators for even static black holes in $AdS_{>3}$ can only be computed numerically\footnote{Note that the full form of the propagators might not be needed to arrive at the conclusion.}.  
\\\\
The fact that rotating black-holes in $AdS_3$ see a different Lyapunov index than the temperature suggests that the bound on chaos \cite{Maldacena:2015waa} computed for thermal large-$N$ systems may be modified for systems with certain chemical potentials turned on. It was shown by Halder  \cite{Halder:2019ric} that by carefully repeating the arguments of \cite{Maldacena:2015waa} for a large-$N$ thermal system with a chemical potential $\mu$ the bound can be derived to be saturated at 
\be
\lambda_L\le \frac{2\pi}{\beta(1-\mu)}
\ee       
provided there are enough states charged under the symmetry for which the chemical potential is turned on. This form of analysis crucially depends on analysing how analytic stucture of the 4pt. function for $\mathcal{O}(1)$ operators  can be restricted in the complex time plane. Shorter the region of analyticity in the Euclidean time, greater is the Lyapunov index.
\\\\
Black holes classically are characterized by their mass, angular momentum and electromagnetic charge $\{M,J,Q\}$. As turning on $J$ and/or $Q$ amounts to turning on a chemical potential in the dual theory, we may expect that the 3-dim results summarised above might even hold for at least rotating black holes in $AdS_{>3}$. This would also seem to be consistent with the result of \cite{Halder:2019ric} and the fast scrambling conjecture which states that black holes are amongst the fastest scramblers of information in the universe.
\section{JT analysis for near extremal BTZ}
The JT action for any geometry close to extremality is obtained by {\bf a)} dimensionally reducing the full gravity action along non-radial space-like directions to obtain a dilaton $\Phi$ coupling to $\sqrt{g}R$ in 2dim without a kinetic term for the former\footnote{The 2d metric has to be scaled appropriately with the dilaton to achieve this. }. {\bf b)} By expanding the 2dim action upto linear fluctuations of the dilaton over it's extremal value $\Phi=\Phi_{\rm ext}+\psi$. Note: before doing so all other fields resulting from the dimensional reduction are to be determined w.r.t. the dilaton $\Phi$. 
\subsection{Thermodynamics of JT model}
We carry out this exercise for the case of gravity in $AdS_3$ which has the following action
\begin{equation}
S_{(3)}=-\frac{1}{2\kappa}\int dx^3 \sqrt{-g}(R-2\Lambda)-\frac{1}{\kappa}\int_\partial dx^2\sqrt{-h}\left(K+\frac{1}{l}\right) 
\label{BTZ_action}
\end{equation} 
where $\kappa=8\pi G_N$.
For dimensional reduction about the non-radial space-like coordinate $y$ we choose the following form of the metric: 
\begin{equation}
ds^2=\Phi^{2\alpha}(ds_{(2)}^2)+\Phi^2(dy+A_t dt)^2,
\label{dim_red_metric}
\end{equation}  
with $\alpha$ to be determined such that the resulting action does not have a kinetic term for the dilaton $\Phi$\footnote{It turns out that for 3-dim one can choose any non-negative value of $\alpha$.}. For the above form of the metric the bulk Lagrangian reduces to
\begin{equation}
\sqrt{-g}( R-2\Lambda)= \sqrt{-\bar{g}}\,\Phi\left(\bar{R}-2\Lambda+\frac{1}{4}\Phi^2 F^2\right),\hspace{0.5cm}{\rm for}\,\,\,\alpha=0
\label{3to2_action}
\end{equation}
where we have chosen $\alpha=0$ for later convenience. All barred quantities are used to denote quantities computed in 2-dim.
Next we would like to solve for $F_{\mu\nu}$ in terms of the dilaton $\Phi$
\be
\nabla_\mu (\Phi^3 F^{\mu\nu})=0 
\label{F_eom}
\ee
which for any 2dim metric takes the form 
\be
F_{rt}=\sqrt{-\bar{g}}\frac{Q}{\Phi^3}
\label{F_on_shell}
\ee
where $Q$ is to be determined further by the dilaton $e.o.m.$\footnote{Note that this is not the most generic solution but is enough for our purpose as we will be expanding about stationary solutions.} Therefore the action now looks like
\be
S_{(2)}=-\frac{1}{2\kappa}\int \sqrt{-\bar{g}}\Phi\left(\bar{R}-2\Lambda-\frac{Q^2}{2\Phi^4}\right)-\frac{1}{\kappa}\int \sqrt{\bar{\gamma}}\Phi \bar{K}. 
\label{2dim_action}
\ee
We choose to separately add the holographic counter terms in the near horizon region and do not keep track of them while dimensional reduction. Next we look at the near horizon metric in the extremal BTZ case given by \eqref{BTZ_nhext} 
\be
\frac{ds^2}{l^2}=\frac{1}{4}\left[\frac{dr^2}{r^2}-r^2d\tau^2\right]+r_+^2\left(d\phi+\frac{r}{2r_+}d\tau\right)^2
\label{BTZ_nhext_appendix}
\ee
where the value of the dilaton $\Phi=\Phi_{\rm ext}=r_+$ which is basically the radius of the sphere at the horizon. Here the gauge field $A_\tau=r/(2r_+)$ fixes the value of the charge $Q$ in  \eqref{F_on_shell} as the above metric is on-shell with regards to the action $S_{(2)}$ \eqref{2dim_action}. Linearizing the dilaton as $\Phi=r_++\psi$ and expanding to linear powers in $\psi$ we get 
\be
S_{\rm JT}=S_{(0)} -\frac{1}{2\kappa}\int \sqrt{-\bar{g}}\,\psi\left(\bar{R}-2\Lambda+\frac{3Q^2}{2r_+^4}\right)-\frac{1}{\kappa}\int_\partial\sqrt{-\bar{\gamma}}\,\psi\, 
\bar{K}
\label{JT_action_appendix}
\ee
where $S_{(0)}=S_{(2)}\big\vert_{\rm ext}$ is the topological term evaluated with only the gauge field and the dilaton taking their on-shell values as given in the metric \eqref{BTZ_nhext_appendix}
\be
S_{(0)}=-\frac{1}{2\kappa}\int \sqrt{-\bar{g}}\,r_+\left(\bar{R}-2\Lambda-\frac{Q^2}{2r_+^4}\right)-\frac{1}{\kappa}\int_\partial \sqrt{-\bar{\gamma}}\,r_+
\bar{K}
\label{topological_action_appendix}
\ee
In order that the metric \eqref{BTZ_nhext_appendix} solves the $\psi$ $e.o.m.$ resulting from $S_{\rm JT}$ for $\psi=0$ we see that the charge is given by
\be
Q=2r_+^2
\label{2dim_charge_appendix}
\ee 
Thus re-writing $S_{\rm JT}$ to be
\bea
&&S_{\rm JT}=S_{(0)} -\frac{1}{2\kappa}\int \sqrt{-\bar{g}}\,\psi\left(\bar{R}-2\bar{\Lambda}\right)-\frac{1}{\kappa}\int_\partial\sqrt{-\bar{\gamma}}\,\psi 
\bar{K}+S_{\rm counter}
,\cr&&\cr
&&\bar{\Lambda}=\Lambda-\tfrac{3}{l^2}=-\tfrac{4}{l^2}
\label{JT_action}
\eea
where $l$ is the radius of $AdS_3$ which we have set to unity. Thus the $\psi$ $e.o.m.$ implies $\bar{R}-2\bar{\Lambda}=0$ for on-shell fluctuations of the 2-dim metric in the JT model. 
\\\\
Next we turn on a temperature in the 2dim theory and therefore work with the metric
\be
ds^2_{(2)}=\frac{1}{4}\left[\frac{dr^2}{r^2}-\left(r-\frac{T_H^2}{4r}\right)^2d\tau^2\right]
\label{thermal_AdS2_appendix}
\ee
where $T_H=2\pi/\beta$ and the horizon occurs at $r=\delta r_+=T_H/2$.
The $2^{\rm nd}$ and the $3^{\rm rd}$ terms in the bulk of topological piece $S_{(0)}$ cancel
yielding the euclidean on-shell action to be
\be
S_{(0)}=-\frac{r_+}{2\kappa}\left(\int\sqrt{\bar{g}}\bar{R}+2\int_\partial\sqrt{\bar{\gamma}}K\right)=-\frac{\beta r_+}{2\kappa} \left(2\delta r_++\frac{T_H^2}{2\delta r_+}\right)=-\frac{4\pi r_+}{2\kappa}.
\ee 
Here the Gibbons-Hawking term at the boundary cancels the bulk divergence while the finite contribution comes only from the horizon at $r=\delta r_+$. Thus the topological piece rightly reproduces the BTZ entropy at extremality with the Free energy $F$ given by the thermodynamic relation
\be
\beta F= \beta(M-\mu J)-S_{\rm ent}
\ee
and  $r_+$ denoting the extremal horizon. Here as BTZ is dual to a large-$N$ system with inverse temperature $\beta$ and chemical potential $\beta\mu$\footnote{$\beta=\frac{2\pi r_+}{r_+^2-r_-^2}$ \& $\mu=\frac{r_-}{r_+}$ for a Lorentzian BTZ.} .
The BTZ free energy  close to extremality reads
\bea
\beta F&=& \beta(M-\mu J)-4\pi r_+\cr
&=&-4\pi (r_++\delta r_+)+\beta(M_{\rm ext}-\mu J_{\rm ext})+\beta(\delta M-\mu \delta J)\cr
&=&-4\pi r_++\beta(M_{\rm ext}-\mu J_{\rm ext})-\beta(1+\mu)\delta M 
\label{BTZ_thermo}
\eea
where in the second line we have expanded about the extremal value of the horizon $r_+\rightarrow r_++\delta r_+$ with $r_+$ labelling the extremal horizon. The equality between the second and the third lines holds for small values of $\delta r_+= T_H/4$. We have also expressed  $\delta M$ \& $\delta J$ for small temperatures as\footnote{We have set $2\kappa$ which has $G_N$ to one, this can also be thought of absorbing $2\kappa$ into the definition of $F,M$ \& $J$. }
\be
 M=M_{\rm ext} +2\delta r_+,\hspace{0.5cm} J=J_{\rm ext}-2\delta r_+\hspace{0.3cm}
 \implies \hspace{0.3cm} \delta M=2\delta r_+=-\delta J
\label{delta_BTZ_MJ}
\ee
The last term in \eqref{BTZ_thermo} is reproduced correctly by the boundary term of the JT model proportional to the dilaton $\psi$ $i.e.$ by euclidean on-shell value of $S_{\rm GH}=S_{\rm JT}-S_{(0)}$.
\be
-S_{\rm GH}=\frac{1}{\kappa}\int_\partial\sqrt{\bar{\gamma}}\,\psi K=\frac{1}{2\kappa}\beta \psi_0 \left(2r^2_\infty +\frac{T_H^2}{2}\right)
\label{JT_onshell} 
\ee
where $r_\infty \implies r\rightarrow\infty$ and we have used the euclidean version of the 2dim metric \eqref{thermal_AdS2_appendix} with $\psi\rightarrow r\psi_0$ at the boundary of the $AdS_2$.  The counter term required for the JT action can be determined by allowing local terms that get rid of the above divergence. This is found to be
\be
S_{\rm counter}=\frac{2}{\kappa}\int_\partial \sqrt{\bar{\gamma}}\,\psi=\frac{1}{2\kappa}\beta \psi_0 \left(2r^2_\infty -\frac{T_H^2}{2}\right)
\label{JT_counter_onshell}
\ee 
Therefore the regularized on-shell euclidean value of $S_{\rm JT}-S_{(0)}$ is 
\be
S_{\rm GH}+S_{\rm counter}=-\beta \psi_0 T_H^2=-\beta(1+\mu)\delta M
\label{JT_onshell_reg}
\ee
where we have used $\psi_0=1/4$ and 
\be
\frac{T_H^2}{4}=(1+\mu)\delta M
\ee
which can be seen to hold for small temperatures. Thus the JT model captures the departures to the Free energy $\beta F$ away from extremality. 
\subsection{Thermal modes}
The near horizon limit of \eqref{BTZ_metric_wiki} at extremality can be achieved in 2 ways; 
by taking simultaneously the extremal and near horizon limit \cite{Gupta:2008ki}
\bea
&& y=\phi+\frac{r_-}{r_+}t,\hspace{0.5cm}r_+-r_-=2\epsilon,\hspace{0.5cm}\rho=r_++\epsilon(r-1),\hspace{0.5cm}t=\frac{\tau}{4\epsilon}\cr
&&\frac{ds^2}{l^2}=\frac{1}{4}\left[\frac{dr^2}{(r^2-1)}-(r^2-1)d\tau^2\right]+r_+^2\left(d\phi+\frac{r-1}{2r_+}d\tau\right)^2
\label{BTZ_nhnext}
\eea
or by taking the near horizon limit of extremal BTZ
\bea
&& y=\phi+t,\hspace{0.5cm}\rho=r_++\epsilon r,\hspace{0.5cm}t=\frac{\tau}{4\epsilon}\cr
&&\frac{ds^2}{l^2}=\frac{1}{4}\left[\frac{dr^2}{r^2}-r^2d\tau^2\right]+r_+^2\left(d\phi+\frac{r}{2r_+}d\tau\right)^2
\label{BTZ_nhext}
\eea
In the JT analysis the above box bracket denotes the Lorentzian analogue of the Euclidean $AdS_2$ disk.   
We would choose to work with the latter geometry.
Different $AdS_2$ geometries can either be denoted by different boundaries of this disk as visualised in \cite{Maldacena:2016upp}, or equivalently by performing PBH diffeomorphisms on the metric while keeping the boundary fixed. In (and only in) 2-dim these are equivalent, we will be adopting the latter of the 2 descriptions as it can be readily generalized to higher dimensions\footnote{In $AdS_3$ these are indeed the diffeomorphisms discussed in section-2, in higher than 3-dim PBH diffeos would correspond to the finite dimensional conformal algebra of the boundary. }.
\\\\
From the 2-dim point of view one can readily read off the the value of the dilaton and the gauge field at extremality from either \eqref{BTZ_nhext} (or \eqref{BTZ_nhnext}) to be $\Phi_{\rm ext}=r_+^2$ and $A_\tau=\frac{r}{2r_+}$.
Note that in the case of near horizon geometries of extremal black holes the dilaton is always independent of the near-horizon radial direction as it is a constant signifying the $S^{d-2}$ sphere volume.
Since \eqref{BTZ_metric_wiki} depends only on radial direction in its components, on-shell JT configurations correspond to on-shell configurations $w.r.t$ the 3-dim action too, as this is a consistent truncation.
\\\\
The $e.o.m.$ for $\psi$ in \eqref{JT_action} allows the 2-dim metric to be locally $AdS_2$ with $\bar{\Lambda}$ being determined by the 3-dim cosmological constant and $\Lambda$ \& $r_+$ as given in \eqref{JT_action_appendix}. The 2-dim metric in box brackets \eqref{BTZ_nhext} therefore can be promoted to a famliy of solutions parametrized by conformal transformations of the coordinate $\tau\rightarrow f(\tau)$ at the throat boundary. 
\begin{equation}
ds^2_{(2)}\equiv\frac{dr^2}{r^2}-r^2d\tau^2+\frac{\{f,\tau\}}{2}d\tau^2-\frac{\{f,t\}^2}{16r^2}d\tau^2
\label{0temp_ads2_fmaily}
\end{equation}
These solutions are obtained by knowing the finite diffeomorphisms (see Appendix A) that correspond to $\tau\rightarrow f(\tau)$ at the conformal boundary.
The on-shell action \eqref{JT_action} then evaluates to the Schwarzian of the conformal transformation
\be
\{f,t\}=-2\,{\rm Sch}[f(t),t]=\frac{3{f''}^2-2f'f'''}{{f'}^2} ,
\label{Schwarzian_def}
\ee
at the boundary of near horizon throat region. If one then uses this action to compute the corrections to the scalar primary 4-pt functions (dual  to 2 bulk scalars minimally coupled to gravity) then one would find $\lambda_L=0$, as one should expect for a black hole at extremality.
\\\\
In order to get an answer for small temperatures the coordinate $\tau$ is made periodic in Euclidean time with period $\beta=T^{-1}_H$. In terms of the PBH diffeomorphisms this amounts to 
\begin{equation}
\left[\frac{dr^2}{r^2}-r^2d\tau^2\right]\rightarrow\left[\frac{dr^2}{r^2}-r^2d\tau^2+\frac{(2\pi T_H)^2}{2}d\tau^2-\frac{(2\pi T_H)^4}{16r^2}d\tau^2\right]=\left[d\rho^2-\sinh^2\!\rho\,\, (2\pi T_H)^2d\tau^2\right]
\label{temp_ads2}
\end{equation}
The similar family of $AdS_2$ metric parametrized $f(\tau)$ now looks like:
\begin{equation}
ds^2_{(2)}\equiv\frac{dr^2}{r^2}-r^2d\tau^2+\frac{\{f,\tau\}+(2\pi T_H)^2f'^2}{2}d\tau^2-\frac{(\left\{f,t\}+(2\pi T_H)^2f'^2\right)^2}{16r^2}d\tau^2
\label{temp_ads2_family}
\end{equation}
thus yielding $\lambda_L=2\pi T_H$ as shown in \cite{Maldacena:2016upp}. Although the value of the dilaton $\psi$ is unimportant for computing the effective action it is important to know that it is constrained by the metric $e.o.m.$ arising from $S_{\rm JT}$ 
\begin{equation}
\nabla_a\nabla_b \psi +g_{a,b}(1-\nabla^2)\psi=0\implies \psi=r\alpha
\label{dilatom_eq}
\end{equation} 
for time independent configurations of the dilaton. It is this divergent value of the linear fluctuation of the dilaton at the throat boundary that signals the departure from the extremal configuration.  An important point to note is that the above equation does not allow for $\psi=const.\neq0$ as a solution.  
\\\\
The fact that this description is thermodynamically consistent for describing near extremal black holes can be seen by computing $S_{\rm JT}$ in \eqref{JT_action} and $S_{(3)}$ in \eqref{BTZ_action} where for the JT action we use the 2-dim geometry \eqref{temp_ads2} at small but finite temperature  along with \eqref{dilatom_eq} with $\alpha=1/4$. 
\\\\
We would like to understand better the effect of the diffeomorphism \eqref{temp_ads2} in terms of the 3-dim throat geometry. We note that the metric \eqref{BTZ_nhext} simply written as
\begin{equation}
ds^2=\frac{dr^2}{4r^2}+r r_+d\tau d\phi +r^2_+d\phi^2
\end{equation}
can be brought to the form
\begin{equation}
ds^2=\frac{dr^2}{r^2}+r^2d\tau d\phi+r_+^2d\phi^2+\frac{(2\pi T_H)^2}{4}d\tau^2+\frac{(2\pi T_H)^2r^2_+}{4r^2}d\tau d\phi
\label{nhnext_temp_0}
\end{equation}
via a finite PBH diffeomorphism. This very same diffeomorphism when restricted to the $AdS_2$ part in the box brackets in \eqref{BTZ_nhext} implies \eqref{temp_ads2}\footnote{This is done by keeping all $\phi$ dependencies, the final answer has no terms proportional to $d\phi$.}${^,}$\footnote{Also note that applying the $AdS_2$ diffeos resulting in \eqref{temp_ads2_family} to \eqref{BTZ_nhext} would not yield a 3dim metric with Dirichlet boundary conditions. }. The above metric is exactly similar to the BTZ metric in the previous section in light cone coordinates $\{r,x^+,x^-\}$. Note also that both the non-radial co-ordinates in the above metric are light like at the throat boundary. Therefore the 3-dim metric\eqref{nhnext_temp_0} in the throat region is the equivalent starting point for describing the throat dynamics (like \eqref{BTZ_nhext}). 
\\\\
Next we would like to see the effect of on-shell solutions to $S_{\rm JT}$ on this geometry.
The effect of $S_{\rm JT}-S_{(0)}$ in \eqref{JT_action} is to describe linearized deformations away from the extremality. Thus implying $\Phi=r_++\psi$, with $\psi$ satisfying \eqref{dilatom_eq} for time independent configurations. As $\psi=\alpha r$ solves the $e.o.m.$ we see that in order to describe deviations from extremality \eqref{nhnext_temp_0} becomes 
\begin{equation}
ds^2=\frac{dr^2}{r^2}+r^2d\tau d\phi+(r_+^2+2\psi)d\phi^2+\frac{(2\pi T_H)^2}{4}d\tau^2+\frac{(2\pi T_H)^2(r_+^2+2\psi)}{4r^2}d\tau d\phi + \mathcal{O}(\psi^2)
\label{nhnext_temp_1}
\end{equation}
This must correspond to a diffeomorphism of \eqref{nhnext_temp_0} as all solutions to the  3-dim  gravity system are diffeomorphic to each other. We note that this particular diffeomorphism is not one of the PBH type as the on-shell value of $\psi$ in the $S_{\rm JT}$ theory diverges towards the boundary. This is precisely the regime where the JT formulation is useful. 
\\\\
Next we turn our attention to the family of solutions \eqref{temp_ads2_family} which give rise to the chaotic behaviour with $\lambda_L=2\pi/\beta$. It can be easily seen that these can be generated by the PBH diffeomorphisms of \eqref{nhnext_temp_0} in the 3-dim throat region giving rise to    
\begin{equation}
ds^2\equiv\frac{dr^2}{r^2}+r^2d\tau d\phi+r_+^2d\phi^2+\frac{\{f,\tau\}+(2\pi T_H)^2f'^2}{4}d\tau^2+\frac{(\{f,\tau\}+(2\pi T_H)^2f'^2)r^2_+}{4r^2}d\tau d\phi.
\label{nhnext_temp_0_family}
\end{equation}
where linear deformations away from extremality of the form of \eqref{nhnext_temp_1} have not been captured as the above metric solves the $AdS_3$ $e.o.m.$ exactly to all orders. However one has a very precise description of these modes atleast  where such corrections are not present $i.e.$ at extremality, the metric in the throat region is exactly
\begin{equation}
ds^2=\frac{dr^2}{r^2}+r^2d\tau d\phi+r_+^2d\phi^2+\frac{\{f,\tau\}}{4}d\tau^2+\frac{\{f,\tau\}r^2_+}{4r^2}d\tau d\phi.
\label{nhext_family}
\end{equation}
We would call these modes thermal as they seem to contribute to chaos only at finite temperatures. Note these modes in some sense capture average of the contribution coming from the modes discussed the the previous section, each of which contribute $\lambda^+_L$ and $\lambda^-_L$ as seen from the boundary of BTZ. 
\section{Extremal throat description}
We next turn our attention to the extremal throat region which has \eqref{BTZ_nhext} as the metric. As we are now in an exactly extremal setting the throat describes an $AlAdS_3$. As seen in section 2 the modes that contributed to chaos at extremality are modes that allowed one to transform from one extremal BTZ solution to another. Thus to describe these modes in the throat region we only need look for those modes which change the value of the extremal horizon appropriately. These modes must also  result in $\lambda_L=2r_+$ $i.e.$ must have a temperature $2\pi T_L=2r_+$ as this is what the higher for the Lyapunov exponents would be at extremality. These modes are obtained after first rescaling $r\rightarrow r^2/r_+$ in \eqref{BTZ_nhext} yielding
\begin{equation}
ds^2=\frac{dr^2}{r^2}+r^2d\tau d\phi +r_+^2d\phi^2
\label{BTZ_nhext_0}
\end{equation}
and then applying the finite form of the PBH transformations corresponding to conformal transformations of $\phi\rightarrow g(\phi)$.
\begin{equation}
ds^2=\frac{dr^2}{r^2}+r^2d\tau d\phi +\frac{1}{4}(\{g,\phi\}+4r_+^2g'^2)d\phi^2
\label{BTZ_nhext_ext_family}
\end{equation}
Note the shift in the Schwarzian is precisely of the form that we expected from the analysis of the full BTZ geometry \cite{Poojary:2018eszz}. We label these modes as the extremal modes.
\\\\
As the extremal throat is perfectly captured by \eqref{BTZ_nhext_0} we can indeed analyse the effective gravity action at the throat boundary with Dirichlet boundary condition. This would imply allowing for conformal transformations of the form  $\tau\rightarrow f(\tau)$ too, which give rise to the thermal modes in the JT analysis of the previous section.
\begin{equation}
ds^2=\frac{dr^2}{r^2}+r^2d\tau d\phi +\frac{1}{4}(\{g,\phi\}+4r_+^2g'^2)d\phi^2+\frac{\{f,\tau\}}{4}d\tau^2+\frac{\{f,\tau\}(\{g,\phi\}+4r_+^2g'^2)}{16r^2}d\tau d\phi.
\label{BTX_nhext_tep_0_ext_family}
\end{equation}
The horizon in the above geometry can be calculated by finding the location at which the area of constant $r$ hyper-surface vanishes, yielding $r_h^4=\{f,\tau\}(\{g,\phi\}+4r_+^2g'^2)$. 
The on-shell action \eqref{BTZ_action} which only gets contribution from the horizon yields
\begin{equation}
S_{3}\equiv\frac{l}{32\pi G_N}\int d\tau d\phi \sqrt{\{f,\tau\}(\{g,\phi\}+4r_+^2g'^2)}
\label{BTZ_nhext_action}
\end{equation}
with $\{\tau,\phi\}$ being the null coordinates on the throat boundary. Note that without incorporating the thermal modes the above action would be zero $i.e.$ for $f(\tau)=\tau$. This is because without the thermal modes the horizon exists at $r=0$ as can be seen from  \eqref{BTZ_nhext_ext_family}. Evaluating propagators for the infinitesimal fluctuations of the fields $f$ \& $g$ above are difficult. However one can choose to make the $\tau$ coordinate transformation in the action by replacing $f(\tau)\rightarrow e^{\alpha f(\tau)}$ thus implying $\{f,\tau\}\rightarrow\{f,\tau\}+\alpha^2 f'^2$. The action now becomes
\begin{equation}
S_{3\alpha}\equiv\frac{l}{32\pi G_N}\int d\tau d\phi \sqrt{(\{f,\tau\}+\alpha^2 f'^2)(\{g,\phi\}+4r_+^2g'^2)}
\label{BTZ_nhext_action_alpha}
\end{equation} 
which is similar to the one obtained in \cite{Poojary:2018eszz} at the conformal boundary of BTZ. One can then proceed to compute the propagators for infinitesimal fluctuations of $f$ and $g$.
The 4-pt scalar $OTOC$ can then be computed using the contributions from these $\alpha$-deformed propagators and then taking the limit $\alpha\rightarrow 0$. This analysis is similar to that in \cite{Poojary:2018eszz} done at the conformal boundary of BTZ  with $x^+\rightarrow \tau$ and $x^-\rightarrow -\phi$ but at extremality. Therefore a similar analysis in the throat region would result in $\lambda^-_L=2r^+$ and $\lambda^+_L=0=\lambda_L$ which matches the extremal limit of the Lyapunov indices obtained in \cite{Poojary:2018eszz}. It may be noted that although turning on the parameter $\alpha$ is like turning on a temperature $\beta^{-1}=\alpha$, the action \eqref{BTZ_nhext_action_alpha} is not valid for finite temperatures as the family of extremal modes at finite temperatures is not known in this region.\footnote{We use this deformation for the ease of doing the computation with \eqref{BTZ_nhext_action_alpha} and are justify it by obtaining the right answer at extremality as seen by taking the extremal limit of the Lyapunov indices at the conformal boundary.} 
\\
Also note that as mentioned before $\{\tau,\phi\}$ are null coordinates at the throat boundary. 
The co-moving time and space coordinates would be given by $\tau=\bar{\tau}+\bar{\phi}$ and $\phi=\bar{\tau}-\bar{\phi}$. 

In order to show that the chaos measured by a $CFT_2$ computation done holographically at the boundary of the throat region we would have to relate the boundary coordinates of the extremal throat $\{\tau,\phi\}$ to the coordinates at the conformal boundary of BTZ $i.e.$ $\{x^+,x^-\}$ in \eqref{Banados_metric}. Note that in taking the near horizon limit in \eqref{BTZ_nhnext} or \eqref{BTZ_nhext} we simultaneously go close to the (boundary of the) horizon throat and to a co-moving coordinate. Therefore these must be related by a diffeomorphism for finite values of $\epsilon$. Further as diffeomorphisms in $AdS_3$ can take one solution to the Einstein's equation to another, these must be "small" diffemorphisms. 

\subsection{Relating boundary and near horizon coordinates}
The extremal throat metric \eqref{BTZ_nhext} is 
\be
\frac{ds^2}{l^2}=\frac{1}{4}\left[\frac{dr^2}{r^2}-r^2d\tau^2\right]+r_+^2\left(d\phi+\frac{r}{2r_+}d\tau\right)^2
\label{BTZ_nhext_appendix}
\ee
Let us re-derive the throat metric and the action \eqref{BTZ_nhext_action} for extremal BTZ in terms of light-cone coordinates $\{x^+,x^-\}$ at the conformal boundary of BTZ.  Extremal BTZ \eqref{BTZ_metric_wiki} in these co-ordinates is simply
\be
\frac{ds^2}{l^2}=\frac{\rho^2d\rho^2}{(\rho^2-r_+^2)^2}-(\rho^2-r_+^2)dx^+dx^-+r_+^2{dx^-}^2.
\label{BTZ_extreme_wiki_metric_appendix}
\ee
The near horizon limits taken in \eqref{BTZ_nhext} in light-cone coordinates read
\be
\rho=r_++\epsilon r,\hspace{0.5cm} x^+=\frac{\tau}{2\epsilon}-x^-.
\label{Sen_limits_extremal_lc_appendix}
\ee
which yields \eqref{BTZ_nhext} as $\epsilon\rightarrow 0$. However since we want to associate the near horizon co-ordinates to the boundary co-ordinates we perform  the following `small' diffeomorphism\footnote{These are `small' as $x^+\rightarrow x^+$ at the conformal boundary of BTZ.}
\bea
&&\rho=r_++\epsilon r,\hspace{0.5cm}x^-\rightarrow x^-,\cr&&\cr
&& x^+\rightarrow x^+-\left[x^+\left(1-\frac{1}{2\epsilon}\right)+x^-\right]\left[\frac{r_+(1+\epsilon)}{r}\right]^{\rm n},\,\,{\rm where}\,\,{\rm n}\in \mathbb{Z}_{>0}
\label{Sen_limits_diffeo_appendix}
\eea
on the metric \eqref{BTZ_extreme_wiki_metric_appendix}. Taking the $\epsilon\rightarrow 0$ limit further, yields
\be
\frac{ds^2}{l^2}=\frac{dr^2}{r^2}-r^2dx^+dx^-+r_+^2{dx^-}^2
\label{BTZ_nhnext_1_lc_appendix}
\ee
which is the near horizon metric \eqref{BTZ_nhext_0} with $\tau\rightarrow x^+$ \& $\phi\rightarrow -x^-$. Therefore the near horizon effective action reads
\be
S_{(3)}=-\frac{1}{2\kappa}\int dx^+ dx^- \sqrt{\{f,x^+\}(\{g,x^-\}+4r_+^2g'^2)}
\label{BTZ_nhext_action_lc_appendix}
\ee
with $f\equiv f(x^+)$ \& $g\equiv g(x^-)$. It is easily seen that as $r\rightarrow\infty$ the transformation \eqref{Sen_limits_diffeo_appendix} takes $x^+\rightarrow x^++\mathcal{O}(r^{-n})$, thus qualifying as a small diffeomorphism.
Note that here it was necessary that we could do a `small' diffeomorphism \eqref{Sen_limits_diffeo_appendix} which essentially allowed us to go to the co-moving co-ordinates on the horizon. 
Now, the late time physics (for $t\gg \beta$) of the CFT$_2$ at the boundary of BTZ is governed by the near horizon region. Therefore the late time dynamics is dictated by the extremal throat $AdS_3/$CFT$_2$. Hence the behaviour of correlators defined on the conformal boundary of the throat region is a good approximation for the behaviour of corresponding correlators on the conformal boundary of BTZ for $t\gg \beta$.
\\\\
Note that for small temperatures the thermal modes described in the previous section capture the average of the 2 temperature inverses as it should $\lambda_L=2\lambda_+\lambda_-/(\lambda_++\lambda_-)$. The modes characterised by $\lambda_+$ in boundary analysis of rotating BTZ are not readily seen in the near horizon region. This is simply the artefact of coordinate transformations the boundary coordinates undergo in order to describe the near horizon geometry as we explain below.  
\\\\
The fact that modes outside the horizon see a different temperature at extremality can also be discerned by computing what is known as the Frolov-Thorne temperature for modes in the throat region. For a generic BTZ metric we can expand quantum fields using boundary coordinates $\{t,y\}$ in terms of asymptotic eigenstates with eigenvalues of energy $\omega$ and angular momentum $m$. This would be a series using basis $e^{-i\omega t+im y}$, or along light-cone co-ordinates using $e^{-in_+x^+-in_-x^-}$. Knowing the co-ordinate transformation that describes the extremal throat region in co-moving co-ordinates $\{\tau,\phi\}$ we can write
\bea
&&e^{-i\omega t+im y}=e^{-in_+x^+-in_-x^-}=e^{-i n_{\rm R} \tau+i n_{\rm L} \phi},\hspace{0.5cm}\cr&&\cr
&&{\rm where}\hspace{0.5cm}t=\frac{\tau}{4\epsilon},\hspace{0.5cm}y=\phi+\frac{\tau}{4\epsilon},\hspace{0.5cm} x^-=-\phi,\hspace{0.5cm}x^+=\frac{\tau}{2\epsilon}+\phi\cr&&\cr
&&\implies  \omega = 4\epsilon\,n_{\rm R}+n_{\rm L},\hspace{0.5cm}m=n_{\rm L},\hspace{0.5cm}n_+=2\epsilon n_R,\hspace{0.5cm}n_-=n_L+2\epsilon n_R
\label{FT_0}
\eea  
where $\epsilon$ is the parameter in the coordinate transformation \eqref{BTZ_nhext}. We then define the left and right temperatures $w.r.t.$ $T_H$ of the black hole as
\bea
&&e^{-(\omega-m \mu)/T_H}=e^{-n_+/T_+-n_-/T_-}=e^{-n_{\rm R}/T_R-n_{\rm L}/T_L}\cr&&\cr
{\rm thus}&&\implies T_L=T_-=\frac{T_H}{1-\mu},\hspace{0.5cm}T_R=\frac{1}{2\epsilon}\left(\frac{T_+T_-}{T_++T_-}\right)=\frac{T_H}{4\epsilon} 
\label{FT_1}
\eea
where $2\pi T_H=\frac{r_+^2-r_-^2}{r_+}=2\left(\frac{T_+T_-}{T_++T_-}\right)$ and $\mu=\frac{r_-}{r_+}$. Upon taking the extremal limit we get
\be
2\pi T_L=2r_+,\hspace{0.5cm} 2\pi T_R=0
\label{FT_temp}
\ee
This is consistent with the microscopic entropy one obtains from the Cardy formula for a unitary CFT with central charge $c_L=3l/2G$. This central charge can also be independently obtained from the throat region from an analysis identical to the one done for the full BTZ as the metric \eqref{BTZ_nhext_1} in the extremal throat region is identical to the extreme BTZ. 
\\\\
We end this section by noting that the extremal modes contribute to chaos only when the thermal modes are also considered. The above action captures their contribution at extremality, it would be interesting to have a prescription to capture their contribution away from extremality for small temperatures as was the case for the thermal modes being described by the JT action.     

\section{Probes in BTZ}

In this section, we will take a different approach. To understand dynamical features of the bulk geometry, we will consider the dynamics of a probe field propagating in this geometry. Towards that, specially in the context of capturing the growth behaviour of OTOCs, a probe string plays a rather crucial role \cite{deBoer:2017xdk}. Not only the probe sector captures salient features of the background geometry, it offers a richer physics, see {\it e.g.}~\cite{Kundu:2015qda,Kundu:2018sof, Kundu:2019ull}. Motivated by this, we will explore the worldsheet temperature of a string probe, assuming implicitly that the corresponding Lyapunov exponent is maximal with the worldsheet temperature. We emphasize that this assumption is fairly mild and should hold on kinematic grounds.

Let us recall the rotating BTZ geometry, with the curvature scale set to unity:
\begin{eqnarray}
ds^2 & = &  - f(r) dt^2 + \frac{dr^2}{f(r)} + r^2 \left( d\phi - \frac{r_+ r_-}{r^2} dt \right)^2 \ , \label{met1} \\
f(r) & = & \frac{\left( r^2 - r_+^2\right) \left( r^2 - r_-^2\right)}{r^2} \ ,  \label{met2}
\end{eqnarray}
where $r_+$ and $r_-$ are the two black hole radii. The mass and the angular momentum of the geometry are given by
\begin{eqnarray}
M = r_+^2 + r_-^2 \ , \quad J = 2 r_+ r_- \ ,
\end{eqnarray}
with the corresponding Hawking temperature and Bekenstein-Hawking entropy:
\begin{eqnarray}
&& T_H = \frac{r_+^2 - r_-^2}{2\pi r_+} \ , \quad \Omega = \frac{r_-}{r_+} \ , \\
&& S_{BH} = \frac{2\pi r_+}{4 G_N} = 2\pi \left( \sqrt{\frac{1}{8 G_N} \left( M + J \right) } + \sqrt{\frac{1}{8 G_N} \left( M - J \right) } \right) \ .
\end{eqnarray}
Here $G_N$ is the Newton's constant and $\Omega$ is the angular velocity at the event horizon. The Hawking temperature can be read off from the standard Euclideanization and subsequently demanding regularity of the Euclidean section. Note that, in the coordinate patch (\ref{met1}), the conformal boundary is represented by an inertial frame, {\it i.e.}~there are no off-diagonal term in the conformal boundary metric.\footnote{Explicitly, this takes the form:
\begin{eqnarray}
ds_{\rm bdry}^{2} = r^2 \left( - dt^2 + d\phi^2 \right)  \ .
\end{eqnarray}
}

Now, we wish to study the dynamics of a probe fundamental string in the background (\ref{met1})-(\ref{met2}). Our goal is to capture the dynamics of a probe degree of freedom in the dual CFT, described by a trajectory $\phi_{{\rm bdry}}(t)$. In static gauge, we choose the following configuration to describe the classical embedding of the string:
\begin{eqnarray}
\tau = t  \ , \quad \sigma = r \ , \quad \phi(t, r ) = \phi_{\rm bdry}(t) + \phi(r) \ , \quad \phi_{\rm bdry}(t) = \omega t \ , \label{anstring}
\end{eqnarray}
where $\omega$ is the angular velocity of the probe particle at the boundary. Clearly, causality constrains $|\omega| \le 1$.\footnote{This is certainly true in the inertial frame; however, not in a non-inertial {\it e.g.}~rotating frame. We will explicitly consider this momentarily.} The dynamics of the string is described by the Nambu-Goto action:
\begin{eqnarray}
&&S_{NG} = - \frac{1}{2\pi \alpha'} \int d\tau d\sigma \sqrt{- {\rm det} \gamma} \ , \\
&& \gamma = \varphi_* [G] \ ,
\end{eqnarray}
where $\alpha'$ is sets the inverse string tension, $\gamma$ denotes the induced worldsheet metric, $G$ is the background metric and $\varphi_*$ is the pull-back map.

The Lagrangian is independent of $\phi(r)$, which readily provides an integral of motion: $(\partial \cL) / (\partial \phi'(r)) = C $, where $C$ is a constant. The embedding profile $\phi'(r)$ has an algebraic solution which can be written as follows:
\begin{eqnarray}
&& \phi'(r) = C \sqrt{\frac{G(r)}{H(r) \left( H(r) - C^2 \right) }} \ , \\
&& G(r) = -\frac{r^2 \left(r^2 \left(\omega ^2 - 1\right) +r_-^2 - 2 r_- r_+ \omega + r_+^2\right)}{\left(r^2 -r_-^2\right) \left(r^2 - r_+^2\right)}\ , \\
&& H(r) = (r -r_-) (r + r_-) (r - r_+) (r + r_+) \ .
\end{eqnarray}
It is clear from the solution above that the point where $G(r)=0$ needs special care. Let us denote this location by $r_{\rm ws}$. For example, at the bulk event horizon this can happen without any issue. For any other point, $r_{\rm ws} > r_+$, this posses an issue: The string cannot end at an ordinary location like $r_{\rm ws}$ in the bulk; this would cause a violation of the charge conservation carried by the end point of the string. However, when $r_{\rm ws} = r_+$, this end point lies inside the black hole and conceptually we are fine, since, classically, it belongs to a causally inaccessible region.

Now, we obtain:
\begin{eqnarray}
G(r) = 0 \quad \implies \quad r_{\rm ws} = \sqrt{\frac{r_+^2 + r_-^2 - 2 r_+ r_- \omega}{1- \omega^2}}  > r_+ \ , \label{rws}
\end{eqnarray}
Thus, we need to extend the string beyond $r_{\rm ws}$ up to $r_+$. This can be implemented by imposing:
\begin{eqnarray}
C = H(r_{\rm ws}) = \frac{1}{1-\omega^2} \left( r_+ - r_- \omega \right)  \left( r_- - r_+ \omega \right)  \ .
\end{eqnarray}
Taking these into account, the solution for $\phi$ can be given by\footnote{Note that, this method is rather generic in probe calculations, see {\it e.g.}~\cite{Karch:2007pd, Albash:2007bq, Alam:2012fw} for a similar behaviour on probe branes.}
\begin{eqnarray}
\phi'(r) = \pm \frac{C r}{\left( r^2 - r_-^2 \right) \left( r^2 - r_+^2  \right)} \frac{\sqrt{r_+^2 + r_-^2 - 2 r_+ r_- \omega - r^2 \left( 1 - \omega^2 \right) }}{ \sqrt{C^2 - \left( r^2 - r_-^2 \right) \left( r^2 - r_+^2  \right) }} \ . 
\end{eqnarray}
The $\pm$ sign contains ambiguity of whether the energy flux is ingoing or outgoing at the event horizon. With these, one can easily write down the induced worldsheet metric, which takes the following schematic form:
\begin{eqnarray}
ds^2 = \gamma_{tt} dt^2 + \gamma_{rr} dr^2 + 2 \gamma_{tr} dt dr \ , 
\end{eqnarray}
which is no longer a diagonal metric. The two-dimensional metric can be easily diagonalized by defining:
\begin{eqnarray}
dt = dt' + h'(r)  dr \ , \quad {\rm with } \quad h'(r) = - \frac{\gamma_{tr}}{\gamma_{tt}} \ ,
\end{eqnarray}
that yields:
\begin{eqnarray}
ds^2 = \gamma_{tt} dt'^2 + \left( \gamma_{rr} - \frac{ \gamma_{tr}^2}{\gamma_{tt} } \right) dr^2 \ , \label{diagws} 
\end{eqnarray}
The worldsheet event horizon, which is found by looking at the zeroes of the inverse of the coefficient of $dr^2$; let us denote this by $r_{\rm ws}^*$. Note that, there is no {\it a priori} reason that $r_{\rm ws}^* = r_{\rm ws}$, where $r_{\rm ws}$ is defined in eqn (\ref{rws}).

Let us find the roots explicitly. The algebraic solution is given by
\begin{eqnarray}
 && \left. \left( \gamma_{rr} - \frac{ \gamma_{tr}^2}{\gamma_{tt} } \right)^{-1} \right |_{r_{\rm ws}^*}= 0 \\
&& \implies \quad  {r_{\rm ws}^*} = \sqrt{\frac{r_+^2 + r_-^2 - 2 r_+ r_- \omega}{1-\omega^2}} \ , \sqrt{\frac{\omega  \left(2 r_- r_+ - \omega \left(r_-^2 + r_+^2\right)\right)}{1-\omega ^2}} \ ,
\end{eqnarray}
while the solution to 
\begin{eqnarray}
\gamma_{tt} = 0 \quad \implies r = \sqrt{\frac{r_+^2 + r_-^2 - 2 r_+ r_- \omega}{1-\omega^2}} = r_{\rm ws} \ .
\end{eqnarray}
Now, it is easy to check that $r_{\pm} < r_{\rm ws}$, as well as ${\rm max}\left( r_{\rm ws}^* \right) = r_{\rm ws}$.

Let us look at the worldsheet geometry near $r=r_{\rm ws}$. The Lorentzian metric in (\ref{diagws}) can now be expanded near the worldsheet event horizon. This yields:
\begin{eqnarray}
ds^2 \approx  - 2 r_{\rm ws}^2 \left( 1 - \omega^2 \right) x^2 dt'^2 + \frac{2 r_{\rm ws}^2}{r_{\rm ws}^2 - r_0^2 } dx^2  \ , \quad r = r_{\rm ws} \left( 1 + x^2 \right) \ ,
\end{eqnarray}
where $r_0$ is a constant determined in terms of $\omega$, and $r_{\pm}$ which we do not explicitly write down. Now, the Euclideanization of the near horizon metric becomes straightforward, by sending $t' \to - i \tau $. Subsequently imposing the periodicity condition, the corresponding temperature is given by
\begin{eqnarray}
&& T_{\rm ws}(\omega, r_+, r_-)^2 = \frac{1}{2} \left [ T_R^2 (\omega +1)^2  + T_L^2 ( 1 - \omega )^2 \right ] \ , \label{temp1} \\
&& T_{R} = \frac{r_+ - r_-}{2 \pi} \ , \quad T_L = \frac{r_+ + r_-}{2 \pi} \ . \label{temp2}
\end{eqnarray}
First of all, the above expression is invariant under:
\begin{eqnarray}
T_L^{\rm ws} \leftrightarrow T_R^{\rm ws} \ , \quad \omega \leftrightarrow - \omega \ .
\end{eqnarray}
This is clearly indicative of the equivalence between the left-movers and the right-movers. Furthermore, below, we enlist some interesting cases in which $T_{\rm ws}(\omega, r_+, r_-)$ takes particularly illuminating forms:
\begin{eqnarray}
&& T_{\rm ws} = \frac{r_+^2 - r_-^2}{2\pi r_+} \ , \quad \omega = \frac{r_-}{r_+} \ , \\
&& T_{\rm ws} = \frac{r_+ - r_-}{\sqrt{2} \pi} = \sqrt{2} T_R = T_{R}^{\rm ws}\ , \quad \omega = 1 \ , \label{twsr} \\
&& T_{\rm ws} = \frac{r_+ + r_-}{\sqrt{2} \pi} = \sqrt{2} T_L = T_{L}^{\rm ws} \ , \quad \omega =  - 1 \ . \label{twsl}
\end{eqnarray}
Clearly, left-movers, right-movers observe the Frolov-Thorne temperatures, up to an overall numerical constant. Meanwhile, a particular linear combination of the left-movers and right-movers observe the Hawking temperature of the background spacetime. This combination knows precisely the angular velocity of the event-horizon. Any other observer would detect a $T_{\rm ws}(\omega)$, for a given value of $\omega$.

Note further that, in the extremal geometry when we set $r_+ = r_-$, the right-moving temperature $T_R = 0$, as well as $T_H =0$. In this case, the worldsheet temperature is observed to be:
\begin{eqnarray}
T = \frac{1}{\sqrt{2}} T_L \left( 1 - \omega \right) \ .
\end{eqnarray}
Thus, for an arbitrary probe degree of freedom, as long as $|\omega| < 1$, the worldsheet fluctuations will couple naturally with $T_L$, where the proportionality constant depends on the angular velocity of the rotating particle. At precisely $\omega=1$, the worldsheet temperature also vanishes since the coupling between the Frolov-Thorne temperature and the worldsheet modes disappear.


Thus, the generic features can be summarized as follows: The worldsheet geometry comes equipped with two independent temperatures: $T_L$ and $T_R$; and the additional parameter of angular velocity at the end point of the string. On a generic point in the parameter space, the worldsheet temperature is a function of all three parameters, as shown in eqn (\ref{temp1}). At special points, {\it e.g.}~at extremality $r_+ = r_-$, the worldsheet modes couple to only one non-vanishing temperature, where the coupling depends on the angular velocity. Correspondingly, the four-point OTOC, following similar calculation as in \cite{deBoer:2017xdk}, will yield a Lyapunov exponent; specifically, even at exact extremality, there are modes on the worldsheet which shows a non-trivial exponential growth in their OTOC, determined by the left-moving temperature. Finally, note that, the left-moving and the right-moving temperatures that we have defined in eqn (\ref{temp2}) are exactly the left-moving and right-moving temperature of the CFT dual to the gravitational background, up to a factor of $\sqrt{2}$.

Before leaving this section, let us offer a few comments on the effective low energy description of the Lyapunov-growth at extremality. It can be checked that the worldsheet is not an AdS$_2$ at a generic point in the parameter space $\{r_+, r_-, \omega\}$. In fact, generically, the worldsheet is nowhere AdS$_2$; thus an effective Schwarzian description cannot arise simply, as described in \cite{Banerjee:2018kwy,Banerjee:2018twd}; however, an effective Schwarzian description arises from the AdS$_3$-perspective, as demonstrated in the previous section. Another way of stating the same fact is to declare that the Schwarzian effective action may not capture the chaotic growing modes at extremality.

\section{Extremal Kerr-$AdS_4$}
In this section we analyse how the near horizon region of extremal Kerr-$AdS_4$ sees the thermal modes in the JT model \cite{Moitra:2019bub} and compare it similarly to the modes that may contribute to extremal chaos.
\\\\ 
The gravity action in $AdS_4$ is 
\be
S_{(4)}=-\frac{1}{16\pi G}\int dx^4 \sqrt{-g}(R-2\lambda)-\frac{1}{\pi G}\int_{\partial}dx^2\sqrt{-\gamma}\left(K -\frac{2}{l}-\frac{l}{2}R_{\partial}\right)
\label{AdS4_action}
\ee
where the boundary terms ($R_{\partial}$ being the boundary intrinsic curvature) included above make the on-shell action finite \cite{Henningson:1998ey,Papadimitriou:2005ii}.
This has a rotating black hole solution given in Boyer-Lindquist coordinates by the metric
\bea
&&ds^2=\rho^2\left(\frac{d\hat{r}^2}{\Delta}+\frac{d\theta^2}{\Delta_\theta}\right)+\frac{\Delta_\theta \sin^2\theta}{\rho^2}\left(ad\hat{t}-\frac{\hat{r}^2+a^2}{\Xi}d\hat{\phi}\right)^2-\frac{\Delta}{\rho^2}\left(d\hat{t}-\frac{a \sin^2\theta}{\Xi}d\hat{\phi}\right)^2\cr&&\cr
&&\rho^2=\hat{r}^2+a^2\cos^2\theta,\hspace{0.5cm}\Delta=(\hat{r}^2+a^2)\left(1+\frac{\hat{r}^2}{l^2}\right)-2m\hat{r},\cr&&\cr
&&\Delta_\theta=1-\frac{a^2}{l^2}\cos^2\theta,\hspace{0.5cm}\Xi=1-\frac{a^2}{l^2}.
\label{Kerr_full}
\eea
with $R_{\mu\nu}=-(3/l^2)g_{\mu\nu}$. The above metric has 2 real horizons $r_\pm$ which are roots of $\Delta=0$. It is algebraically convenient to express thermodynamic quantities mass $M$ and angular momentum $J_\phi$ in terms of $m,a$ which are constant parameters in the above metric rather than $r_\pm$.
\bea
&&2\pi T_H=\frac{r_+^2 -a^2+r^2_+l^{-2}(3r_+^2+a^2)}{2 r_+(r_+^2+a^2)},\hspace{0.5cm} S_{BH}=\frac{\pi(r_+^2+a^2)}{\Xi}\cr&&\cr
&&M=\frac{m}{\Xi^2},\hspace{0.5cm}J_\phi=\frac{ma}{\Xi^2},\hspace{0.5cm} \Omega_\phi=\frac{a\Xi}{r_+^2+a^2}
\label{Kerr_quant}
\eea
where $S_{BH}$ is the black hole entropy and $\Omega_\phi$ its horizon's angular velocity as measured at the boundary. Also note that the metric \eqref{Kerr_full} asymptotes to an $AdS_4$ in rotating boundary coordinates. In order to obtain the thermodynamically consistent description of it's mass and angular momentum one should compute quantities with boundary coordinates being non-rotating \cite{Papadimitriou:2005ii}. The Kerr solution is completely described by specifying the outer horizon radius $r_+$ and the parameter $a$, all other quantities can be expresed in terms of $r_+$ \& $a$. 
\\\\
The extremal limit corresponds to the having coincident roots $r_\pm=r_0$ to the eq. $\Delta=0$, thus implying $\partial_{\hat{r}}\Delta=0$. It is easier to solve for the extremal value of parameters $m$ \& $a$ in terms of $r_0$ rather than  vice-versa. 
\be
m=\frac{r_0(1+r_0^2l^{-2})^2}{1-r_0^2l^{-2}},\hspace{0.5cm}a^2=\frac{r_0^2(1+3r^2_0l^{-2})}{1-r_0^2l^{-2}}
\label{Kerr_ext_ma}
\ee
We first expand $\Delta$ about extremal root $r_0$ as
\be
\Delta=(\hat{r}-r_0)^2V+\mathcal{O}((\hat{r}-r_0)^3),\hspace{0.5cm}{\rm where}\hspace{0.5cm} V=\frac{1+6r_0^2l^{-2}-3r_0^4l^{-4}}{1-r_0^2l^{-2}}
\label{ext_0}
\ee
Like in the BTZ case we would have to go to a co-moving frame in order to describe the extremal throat region in \eqref{Kerr_full} by defining a scaling parameter $\epsilon$ as
\be
\hat{r}=r_0(1+\epsilon r),\hspace{0.5cm}\hat{t}=\frac{r_0^2+a^2}{\epsilon r_0 V}t,\hspace{0.5cm}\hat{\phi}=\phi+\frac{a\Xi}{r_0^2+a^2}\hat{t}
\label{ext_coord}
\ee
The extremal throat metric is then obtained by taking the limit $\epsilon\rightarrow 0$ yielding \cite{Lu:2008jk,Hartman:2008pb}
\bea
&&ds^2=\frac{\rho_0^2}{V}\left(\frac{dr^2}{r^2}-r^2dt^2+\frac{V}{\Delta_\theta}d\theta^2\right)+\frac{4a^2r_0^2\Delta_\theta\sin^2\theta}{V^2\rho^2_0}\left(rdt+\frac{V(r_0^2+a^2)}{4\,a^2r_0^2\,\Xi}d\phi\right)^2,\cr&&\cr
&&\rho^2_0=r_0^2+a^2\cos^2\theta.
\label{Kerr_ext}
\eea 
with $a$ given in \eqref{Kerr_ext_ma}.
The above metric has a constant negative curvature at any fixed value of $\theta$.
As expected the throat metric \eqref{Kerr_ext} exhibits an $AdS_2$  as seen in the coordinates $\{r,t\}$. \newline\newline
For a specific value of $\theta=\theta_0$ given by
\be
\rho_0^4=\frac{4a^2r_0^2\Delta_\theta\sin^2\theta}{V},
\label{theta0}
\ee
 the geometry is exactly the $AdS_3$ and of the form of \eqref{BTZ_nhext}
\bea
&&\frac{ds^2}{4k}=\frac{1}{4}\left[\frac{dr^2}{r^2}-r^2dt^2\right]+R^2_0\left(d\phi+\frac{r}{2R_0}dt\right)^2,\cr&&\cr
{\rm with}&&k=\frac{\rho_0^2}{V}\big\vert_{\theta=\theta_0},\hspace{0.5cm}R_0=\frac{V(r_0^2+a^2)}{4\,a^2r_0^2\,\Xi}.
\label{Kerr_AdS3}
\eea
The metric \eqref{Kerr_ext} possesses an $SL(2,\mathbb{R})\times {\rm U}(1)$ symmetry with the ${\rm U}(1)$ coming from the relabelling of the azimuthal coordinate $\phi$ \cite{Hartman:2008pb}. The metric \eqref{Kerr_ext} is therefore a warped $AdS_3$ at fixed $\theta$ with a $\theta$-dependent warping factor.
\subsection{Thermal modes \& near extremal chaos}
As seen in the BTZ case, here too the contribution coming from the thermal modes is captured nicely by the JT model for black holes with a small temperature turned on close to extremality. These have been studied \& verified in the 4-dim RN-Kerr and 5-dim RN-Kerr cases to great detail by Moitra $et\,\, al$ \cite{Moitra:2019bub}\footnote{Although \cite{Moitra:2019bub} covers the case of charged rotating black hole in $AdS_4$, the details for the rotating case can be obtained by taking the electric chrge to zero.  }. We briefly review the mechanism here for Kerr-$AdS_4$ without going into the computational details.
\\\\
The near horizon metric \eqref{Kerr_ext} for the extremal Kerr-$AdS_4$ can be also written as:
\bea
&&ds^2=\frac{\rho_0^2}{V}\left(\frac{dr^2}{r^2}-r^2dt^2+\frac{V}{\Delta_\theta}d\theta^2\right)+\frac{\Delta_\theta (r_0^2+a^2)\sin^2\theta}{\Xi\rho^2_0}\Phi_{\rm ext}^2\left(d\phi+\frac{2r_0a\Xi}{V(r_0^2+a^2)}rdt\right)^2\!\!\!,\cr&&
\label{Kerr_ext_dilaton}
\eea 
where $\Phi_{\rm ext}^2=\frac{r_0^2+a^2}{\Xi^2}$.
Note that the dilaton $\Phi_{\rm ext}$ is being identified as volume of the the $S^2$ spanned by $\{\theta,\phi\}$ coordinates $V_{S^2}=4\pi\Phi_{\rm ext}^2$.
The above metric is then re-written for generic values of dilaton $\Phi$ as
\bea
&&ds^2=\frac{\rho_0^2}{V}\left(\frac{\Phi_{\rm ext}}{\Phi} ds^2_{(2)}(r,t)+\frac{V\Phi^2}{\Delta_\theta \Phi_{\rm ext}^2}d\theta^2\right)+\frac{\Delta_\theta(r_0^2+a^2)\sin^2\theta}{\Xi\rho_0^2}\Phi^2\left(d\phi+A_t(r,t)dt\right)^2\!\!\!\!,\cr&&
\label{Kerr_ext_dilaton_1}
\eea
which is motivated by the fact that upon dimensional reduction on to 2-dim the dilaton $\Phi(r,t)$ does not have a kinetic term. The 2-dim action resulting from the dimensional reduction of \eqref{AdS4_action} over the above metric is quite complicated \cite{Moitra:2019bub} apart from having the $\Phi^2 \bar{R}$ part in its Lagrangian. Here $\bar{R}$ is the 2 dimensional Ricci scalar for the resulting geometry spanned by the remaining  $\{r,t\}$ coordinates. Also note that the field strength $F_{rt}$ for gauge field $A_t$ resulting from dimensional reduction can be  solved in terms of $\Phi$ \cite{Moitra:2019bub}. We have demonstrated this for the case of $BTZ$ in \eqref{F_on_shell} in subsection-3.1.
\\\\
Upon dimensionally reducing the action \eqref{AdS4_action} along \eqref{Kerr_ext_dilaton_1} and solving gauge field $e.o.m.$ one obtains an action for the 2-$dim$ metric $ds^2_{(2)}=\bar{g}_{ab}dx^adx^b$ and $\Phi$.
Expanding this action for linear fluctuations of the dilaton as before $\Phi=\Phi_{\rm ext}+\psi$, one arrives at the JT action \eqref{JT_action}. The JT action therefore captures the contribution coming from the different $AdS_2$ geometries given by the metric $\bar{g}_{ab}$ above. As we have turned on a fluctuation $\psi$ over the extremal value of the dilaton, the $AdS_2$ geometry must have a temperature $\beta^{-1}$. Thus the JT action captures the $AdS_2$ fluctuations about a thermal $AdS_2$ given by
\begin{equation}
\bar{g}_{ab}dx^adx^b=\left[\frac{dr^2}{r^2}-r^2d\tau^2+\frac{(2\pi T_H)^2}{2}d\tau^2-\frac{(2\pi T_H)^4}{16r^2}d\tau^2\right]=\left[d\rho^2-\sinh^2\!\rho\,\, (2\pi T_H)^2d\tau^2\right]
\label{temp_ads2_1}
\end{equation} 
The extremal parameters are described by $\Phi_{\rm ext}^2=(r_0^2+a^2)/\Xi$, which basically is a function of extremal horizon $r_0$. Different values of horizon radius therefore specify different extremal solutions.  \newline\newline
Note that $\psi$ satisfies an on-shell equation \eqref{dilatom_eq}
\begin{equation}
\nabla_a\nabla_b \psi +\bar{g}_{a,b}(1-\nabla^2)\psi=0
\label{dilatom_eq_1}
\end{equation} 
which does not allow a  $\psi=const.\neq 0$ as a solution. Therefore although the full-non-linear $e.o.m.$ constraining  $\psi$ must allow for constant shifts in the value of the dilaton, linearization about a constant value $\Phi_{\rm ext}$ given above doesn't. It is this space of extremal solutions which would be important for studying extremal chaos.   
\subsection{Extremal chaos}
In order to study extremal modes that contribute to chaos similar to those in BTZ we look for those 3-dim diffeomorphisms which take the metric \eqref{Kerr_ext} from extremal configuration to another. As these are exactly known in the case of BTZ geometry and therefore for metric \eqref{Kerr_AdS3} we can simply apply them to the throat metric \eqref{Kerr_ext}. However we can only do this in it's infinitesimal form\footnote{We return to this point later.}. Such extremal diffeomorphisms of \eqref{Kerr_AdS3} are generated by vectors of the form
\be
\xi_{\rm ext}^\mu\partial_\mu=-rf'(\phi)\partial_r + f(\phi)\partial_\phi + \mathcal{O}(r^{-2})
\label{Kerr_ext_vir_vector}
\ee  
which are precisely the Brown-Henneaux vector fields in the 3-dim space corresponding to infinitesimal conformal transformations of coordinate $\phi\rightarrow \phi+f(\phi)$ . These are exactly the vector fields studied in the context of Kerr/CFT \& Kerr-AdS/CFT correspondence \cite{Bredberg:2009pv,Hartman:2008pb,Hartman:2009nz,Compere:2012jk}. The throat metric \eqref{Kerr_ext} is allowed to fluctuate to precisely those orders in $r$ in its components as is allowed by the Lie derivative of  metric \eqref{Kerr_ext} along the vector \eqref{Kerr_ext_vir_vector}.
\be
\mathcal{L}_{\xi_{\rm ext}}g_{\mu\nu}=h_{\mu\nu}\approx
\begin{pmatrix}
\mathcal{O}(1/r^3) & \mathcal{O}(1/r^2) & \mathcal{O}(1/r^2) & \mathcal{O}(1/r)\\
\;&\mathcal{O}(r^2) & \mathcal{O}(1/r) & \mathcal{O}(1) \\
\;&\;&\mathcal{O}(1/r) & \mathcal{O}(1/r)\\
\;&\;&\;&\mathcal{O}(1)	 
\end{pmatrix}
\label{Kerr_ext_falloff}
\ee
in $\{r,t,\theta,\phi\}$ coordinates, here $g_{\mu\nu}$ is \eqref{Kerr_ext}. 
\\\\
Like in the BTZ case, above given  fall-off conditions would then allow a set of consistent large diffeomorphisms. Note these are not of the Dirichlet type. It turns out that one  may define another set of diffeomorphisms infinitesimally generated by 
\be
\xi_{t}^\mu\partial_\mu=g(t)\partial_t +\mathcal{O}(r^{-1})
\label{Kerr_ext_thermal_vector}
\ee
along which the Lie derivative of \eqref{Kerr_ext} obeys the above fall-off \eqref{Kerr_ext_falloff}. It is not hard to see that these take the metric away from extremality as relabelling of $t\rightarrow t+g(t)$ allows for a Euclidean conical defect to be generated in the $\{r,t\}$ part of the metric similar to the BTZ case. These are precisely the infinitesimal versions of the diffeomorphisms which are used to generate the thermal $AdS_2$ \eqref{temp_ads2_1} while studying the thermal modes close to extremality\footnote{The thermal $AdS_2$ deformations are infinitesimally generated by $\xi^\mu\partial_\mu=-\frac{r}{2}g'(t)\partial_r+g(t)\partial_t$, but the $r$-component of the vector field can be ignored as the boundary conditions \eqref{Kerr_ext_falloff} are more relaxed than Dirichlet boundary conditions. }. 
Therefore in order to stay extremal only constant shifts in $t$ are allowed. 
\\\\
The space of fluctuations are thus generated infinitesimally by the vector fields
\bea
&&\xi_{\rm ext}^\mu\partial_\mu=-rf'(\phi)\partial_r + f(\phi)\partial_\phi + \mathcal{O}(r^{-2}),\hspace{0.3cm}\xi_{t}^\mu\partial_\mu= \partial_t +\mathcal{O}(r^{-1})
\label{Kerr_ext_vir_t_vectors}
\eea   
These have been studied to a great extent in the context of Kerr/CFT correspondence. These form the asymptotic symmetry algebra of the near horizon extremal Kerr (NHEK) family of geometries. Their asymptotic charges can be defined for the above set of vector fields with regards to the fall-off conditions \eqref{Kerr_ext_falloff} on the spatial slice at the boundary of the throat region. 
\bea
&&Q_{\xi}=\frac{1}{8\pi}\int_{\partial} *\mathcal{K}_{\xi}\cr&&\cr
&&\mathcal{K}_{\xi}[h,g]=\tfrac{1}{2}\left[\xi_\mu\nabla_\mu h-\xi_\mu\nabla_\alpha h^\alpha_{\,\,\nu}+\xi_\alpha\nabla_\mu h_\nu^{\,\,\alpha}+\tfrac{1}{2}h\nabla_\mu\xi_\nu-h_\mu^{\,\,\alpha}\nabla_\alpha\xi_\nu\right.\cr
&&\hspace{4cm}\left.+\tfrac{1}{2}h_\mu\alpha\left(\nabla_\nu\xi^\alpha+\nabla_\alpha\xi_\nu\right)\right]dx^\nu\wedge dx^\mu
\label{charge_0}
\eea
where $*$ denotes the Hodge dual and $\mathcal{K}_\xi$ is the 2-form simplectic current defined on the boundary and $g$ here denotes the throat metric \eqref{Kerr_ext}\footnote{The derivatives in the expression of $\mathcal{K}$ are computed $w.r.t.$ metric $g$.}. These are finite and can be used to define the Poisson structure for the space of solutions allowed by the above conditions.
\\\\
The vectors $\xi_{\rm ext}$s form a Witt algebra for the parametrization $f(\phi)=\sum\xi^{(n)}e^{in\phi}$ under the commutator along their Lie derivatives
\be
\left[ \xi_{\rm ext}^{(m)},\xi_{\rm ext}^{(n)} \right]\equiv\left[\mathcal{L}_{\xi^{(m)}_{\rm ext}},\mathcal{L}_{\xi^{(n)}_{\rm ext}}\right]=i(m-n)\mathcal{L}_{\xi^{(m+n)}_{\rm ext}}
\label{Witt}
\ee
where $\xi_{\rm ext}^{(n)}$ is obtained by replacing $f(\phi)=e^{i n \phi}$ in $\xi_{\rm ext}$. Demanding that the charges associated with the above vectors generate required change in the solution space about the extremal solutions $via$ Lie derivative allows one to define a Poisson bracket on the phase space of allowed solutions.
\be
\{Q_{\xi_{\rm ext}^{(m)}},Q_{\xi_{\rm ext}^{(n)}}\}=Q_{\left[ \xi_{\rm ext}^{(m)},\xi_{\rm ext}^{(n)} \right]}+\frac{1}{8\pi}\int_{\partial} *\mathcal{K}_{\xi_{\rm ext}^{(m)}}[\mathcal{L}_{\xi^{(n)}_{\rm ext}},g]
\label{Charge_algebra}
\ee
 This gives rise to the asymptotic structure of the Virasoro algebra with a central extension given by the last term in \eqref{Charge_algebra}
\be
c_L=12\frac{r_0 a}{V},
\label{c}
\ee 
with $a$ being the extremal value given by \eqref{Kerr_ext_ma}.
\\\\
Like in the BTZ case one can find a temperature associated to these extremal fluctuations. If one where to look at the throat metric at the specific angle $\theta=\theta_0$ given by \eqref{theta0} the metric \eqref{Kerr_AdS3} is exactly $AdS_3$
\bea
&&\frac{ds^2}{4k}=\frac{1}{4}\left[\frac{dr^2}{r^2}-r^2dt^2\right]+R^2_0\left(d\phi+\frac{r}{2R_0}dt\right)^2,\cr&&\cr
{\rm with}&&k=\frac{\rho_0^2}{V}\big\vert_{\theta=\theta_0},\hspace{0.5cm}R_0=\frac{V(r_0^2+a^2)}{4\,a^2r_0^2\,\Xi}.
\label{Kerr_AdS3_1}
\eea
One might be tempted to conclude that since the same metric is obtained in the extremal BTZ near horizon region, a temperature of
\be
2\pi T_L=2R_0=\frac{V(r_0^2+a^2)}{2a^2r_0^2\Xi}=\frac{1+6r_0^2l^{-2}-3r_0^4l^{-4}}{(1-3r_0^2l^{-2})\sqrt{(1-3r_0^2l^{-2})(1-r_0^2l^{-2})}}
\label{Kerr_ext_temp}
\ee
could be expected, with a corresponding $T_R=0$ as the geometry is extremal. 
 One can check the above temperature matches with the Frolov-Throne temperature seen by modes outside the horizon \cite{Lu:2008jk}. The Cardy formula then gives the right extremal entropy with the central charge given by \eqref{charge_0}
\be
S_{\rm ent}=\frac{1}{3}\pi^3c_LT_L=\frac{2\pi r_0^2}{1-3r_0^2l^{-2}}
\label{Kerr_ext_entropy}
\ee
However in the BTZ case we knew the full non-linear completion of the diffeomorphisms corresponding to the extremal modes. This allowed us to ascertain the existence of the Schwarzian function and how it changes if there were a temperature. The full non-linear form of these diffeomorphisms is a privilege to be had only in 2 \& 3 dimensions.   
\\\\
If one were to simply apply the full non-linear transformation which results in \eqref{BTZ_nhext_ext_family} to the extremal throat metric \eqref{Kerr_ext} then it would violate the fall-off conditions \eqref{Kerr_ext_falloff}. This should not be surprising as dimensionally uplifting non-linear diffeomorphisms naively\footnote{With no dependence on the extra dimensions $i.e.$ $\theta$ in this case.} often leads to a metric which violates the fall-off conditions which are respected by the infinitesimal versions of the same diffeomorphisms. This can also be readily seen in the 3-dim case where uplifting the non-linear thermal PBH diffeomorphisms which results in \eqref{0temp_ads2_fmaily} $i.e.$
\begin{equation}
\left[\frac{dr^2}{r^2}-r^2d\tau^2\right]\rightarrow\left[\frac{dr^2}{r^2}-r^2d\tau^2+\frac{\{f,\tau\}}{2}d\tau^2-\frac{\{f,t\}^2}{16r^2}d\tau^2\right]
\label{0temp_ads2_family_1}
\end{equation}
when applied to \eqref{BTZ_nhext} 
\be
\frac{ds^2}{l^2}=\frac{1}{4}\left[\frac{dr^2}{r^2}-r^2d\tau^2\right]+r_+^2\left(d\phi+\frac{r}{2r_+}d\tau\right)^2
\label{BTZ_nhext_1}
\ee
without any $\phi$ dependence  yields a metric which violates the Dirichlet or Brown-Henneaux type boundary conditions. However we know the non-linear diffeomorphisms in 3-dim which do obey the Dirichlet boundary conditions and have the same effect as \eqref{0temp_ads2_family_1} when restricted to 2-dim. 
\\\\
This suggests that there should be some non-linear completion of diffeomorphisms generated by vectors  \eqref{Kerr_ext_vir_t_vectors}  which also depends on $\theta$ such that the resulting metric obeys the fall-offs \eqref{Kerr_ext_falloff}. These have not yet been constructed but would have imporatnt consequences as we will discuss further.
\\\\
Similar to the BTZ case one may proceed to find the effective action in the extremal throat region. Doing so for the family of solutions generated by the asymptotic symmetry generators \eqref{Kerr_ext_vir_t_vectors} would similarly yield an effective action at the throat boundary. This action could in principle be used to study the contribution to the $OTOC$ of  boundary operators dual to minimally coupled bulk scalars. But there are a few technical issues to be addressed first.
\\\\
{\bf{1)}} The fall-offs \eqref{Kerr_ext_falloff} are not the usual Dirichlet boundary conditions one usually imposes in $AdS/CFT$ as the boundary metric is allowed to fluctuate at $\mathcal{O}(r^2)$. This would imply that a necessary boundary counter terms need to be added to make the variational problem $w.r.t.$ the metric well posed.
\\\\
{\bf{2)}} One also needs to write the NHEK metric \eqref{Kerr_ext} and the full space of solutions generated by \eqref{Kerr_ext_vir_t_vectors} about it in the Fefferman-Graham form. This necessary in order to implement holographic renormalization of the on-shell action. For the case of the full Kerr-$AdS_{4,5}$ metrics this was systematically done in \cite{Papadimitriou:2005ii}. The asymptotic charges defined on such a boundary for Killing symmetries are the ones that enter the thermodynamic relations for the black-hole system.
\\\\
{\bf 3)} It would be interesting to find the non-linear completions of both the thermal diffeomorphisms which give rise to the Schwarzian in the JT description, and the extremal diffeomorphisms discussed above for the NHEK geometry such that \eqref{Kerr_ext_falloff} are obeyed. Like the BTZ case it might probably also imply that in order to see the effect of the extremal modes one has to include the contribution coming from the thermal ones\footnote{This might not always be true as in Kerr-$AdS_{4,5}$ the on-shell action does receive finite contributions from the boundary(see \cite{Papadimitriou:2005ii}.), which is not the case in $AdS_3$.}.           
\\\\
We end this subsection by noting that the dynamics of the warped $AdS_3$ in the NHEK metric can contribute to the late time physics of CFT$_3$ correlators at the conformal boundary of Kerr-$AdS_4$. Near horizon considerations similar to that of the BTZ case suggests that there do exist modes that can contribute to extremal chaos. These have been previously studied in the context of Kerr/CFT correspondence and their asymptotic symmetry algebra consists of a Virasoro suggesting the existence of Schwarzian like behaviour. As espoused in the previous sections the extremal mode dynamics are independent of the thermal fluctuations captured by the JT model. The study of these modes is nonetheless important in understanding how black holes scramble information as they do seem to contribute to its chaotic behaviour. It is also worth nothing that extremal configurations in string theory have a good microstate description in terms of brane configurations obeying certain BPS conditions \cite{Lin:2004nb}. It would indeed be very interesting to find how the space for such brane configurations participates in the dynamics of scrambling information. 
\\\\
We next proceed to see what more can a string probe can tell us about extremal temperature as we did in section 5 for the case of extremal BTZ.

\subsection{Probing the Geometry}

As we have done in the BTZ-geometry, let us study the behaviour of a probe string in the AdS$_4$-Kerr background in (\ref{Kerr_full}). First, let us offer some comments. Note that, in the coordinate system of (\ref{Kerr_full}), the conformal boundary metric is written in a rotating frame, with angular velocity $-a$. This can be easily reconciled with the BTZ description, by redefining the boundary coordinates and going to an inertial frame. Towards that, let us note that a generic probe string in this background can be described by the following embedding function: $\{\theta(\hat{r}), \phi(\hat{r}) \}$, in the static gauge: $\tau = \hat{t}$ and $\sigma = \hat{r}$, where $\{\tau, \sigma\}$ are worldsheet coordinates. Now, schematically, the Nambu-Goto action takes the following form:
\begin{eqnarray}
S_{\rm NG} = - \frac{1}{2\pi\alpha'} \int d\tau d\sigma \cL\left[ \theta, \theta', \hat{\phi}' \right] \ .
\end{eqnarray}
Thus, the general equations of motion will yield a set of coupled equations:
\begin{eqnarray}
&& \frac{d}{d\hat{r}} \left( \frac{\partial \cL } {\partial \hat{\phi}'}\right) - \frac{\partial \cL}{\partial \hat{\phi}} = 0 \ , \label{eomphi} \\
&& \frac{d}{d\hat{r}} \left( \frac{\partial \cL } {\partial \theta'}\right) - \frac{\partial \cL}{\partial \theta} = 0 \ . \label{eomtheta}
\end{eqnarray}
It is straightforward to check that the second equation is trivially satisfied at $\theta= \pi/2$, which corresponds to the equatorial embedding of the probe string. Thus, without any loss of generality, we will consider this case. Now, we use the ansatz:
\begin{eqnarray}
\hat{\phi} = \left( \omega - a \right) d\hat{t} + \phi' d\hat{r} \ ,
\end{eqnarray}
where $\omega$ is the angular velocity measured in the inertial conformal boundary frame.

\begin{figure}[t]
\includegraphics[width=10cm]{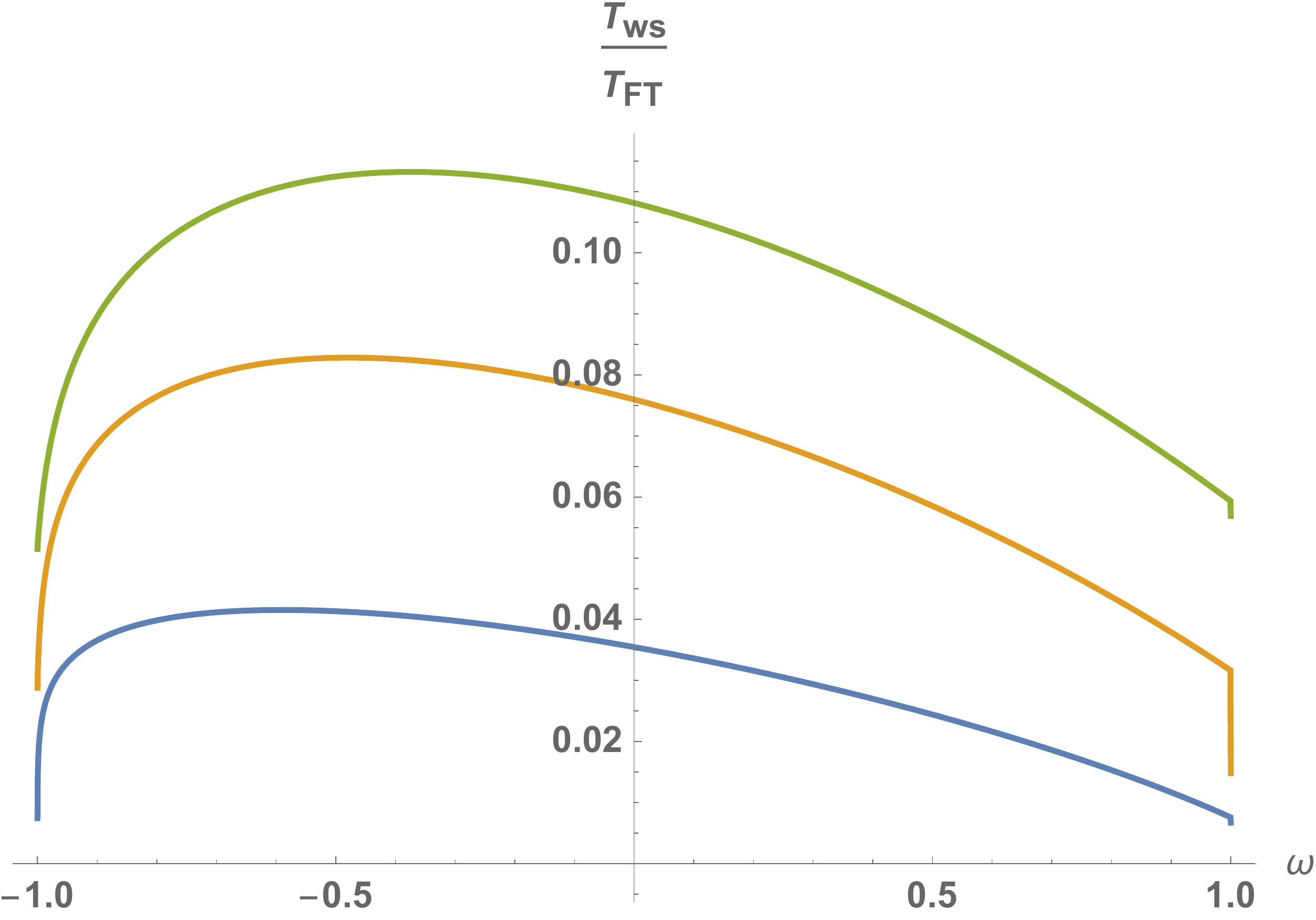}
\centering
\caption{\small We have plotted the worldsheet temperature, in units of the Frolov-Thorne temperature $T_{FT}$, defined in eqn (\ref{Kerr_ext_temp}). Here curves corresponds to setting $r_0 = 1/2, 2/5, 1/3$, from top to bottom, respectively. Clearly, the curves end at $|\omega|=1$, which is the causality bound.} \label{Tws1}
\end{figure}

To proceed further, let us first consider the extremal case. This corresponds to setting:
\begin{eqnarray}
&& m = r_0 \frac{ \left( 1 + r_0^2 \right)^2}{1 - r_0^2} \ , \quad a^2 = r_0^2 \frac{1 + 3 r_0^2}{1 - r_0^2} \ , \\
&& \Delta = \left(\hat{r} - r_0 \right)^2 \left( \hat{r}^2 + 2 \hat{r} r_0 + \frac{1 + 3 r_0^2}{1 - r_0^2 } \right)  \ , \quad {\rm with} \quad 1 > 3 r_0^2 \ , 
\end{eqnarray}
in the geometry in (\ref{Kerr_full}). Now, one proceeds exactly as we have previously done in probing the extremal BTZ geometry. In this case, most of the explicit algebraic expressions are unwieldy and not necessarily illuminating; therefore we do not write them down. At the end of the day, one proceeds to calculate a worldsheet temperature, which is, again, a function of the extremal horizon $r_0$ and the angular velocity of the string end point. This is pictorially demonstrated in figure \ref{Tws1}.

Let us briefly comment on the generic observations: First, note that, the worldsheet temperature now is a function of the Frolov-Thorne temperature\footnote{Note that, this is the analogue of the left-moving temperature at extremality, for BTZ.} and the angular velocity of the string end point. However, unlike in the BTZ case, this functional dependence does not factorize in terms of two independent functions of the extremal temperature (Frolov-Thorne, in this case) and $\omega$. However, the worldsheet extremal limit is set by the causality bound $\omega =1$, which is true for any Frolov-Thorne temperature. Thus, it has a clear similarity to the $T_H \to 0$ and $|\omega| \to 1$ extremality condition in the BTZ background. Note further, given the Frolov-Thorne temperature, the worldsheet temperature has one maxima, as a function of $\omega$, similar to the BTZ case. In the latter, however, because of the ${\mathbb Z}_2$ symmetry: $\omega \to -\omega$, of the worldsheet temperature peaks at $\omega = 1/2$. In higher dimensions, this ${\mathbb Z}_2$-symmetry is broken and the peak occurs asymmetrically along the $\omega$-axis.

Thus, even at extremality, this class of worldsheet fluctuations will yield an OTOC, with an exponentially growing mode with the corresponding Lyapunov exponent, which is now determined by a non-trivial function of the Frolov-Thorne temperature. Once again, the worldsheet is not generically AdS$_2$, and therefore the Schwarzian effective action, as is obtained in the context of JT-gravity, is not relevant here. Generically though, on the worldsheet a notion of the left-movers and the right-movers do remain, but they become chiral.\footnote{This chirality is inherited from the warped AdS$_3$ near horizon near extremal throat, where the dual CFT distinguishes between the left-movers and the right-movers.}

Let us offer some comments on what happens away from extremality. There is no longer any notion of a left-moving and a right-moving temperature in the background. However, as far as the string worldsheet is considered, the corresponding temperature $T_{\rm ws}\left( m, a, \omega\right)$ is likely to have a maximum and a minimum, corresponding to rotation parallel or anti-parallel to the rotation of the event horizon. Away from extremality, $T_{\rm max}^{\rm ws}$ and $T_{\rm min}^{\rm ws}$ will be analogues to $T_{L}^{\rm ws}$ and $T_{R}^{\rm ws}$ in (\ref{twsr})-(\ref{twsl}).

Now it becomes algorithmic to explore higher dimensional AdS-Kerr geometries. For example, take the AdS$_5$-Kerr background. In this case, the boundary metric has an $S^3$ and the corresponding SO$(4)$ has two U$(1)$ Cartans, along which two independent angular velocities can be realized. The simplest profile of the string will be given by the equatorial embedding, where the end point lies on the equator of the $S^3$. At extremality, this background yields two Frolov-Thorne temperatures, see \cite{Lu:2008jk}, and the worldsheet temperature will be a function of these two temperatures, as well as the angular velocity of the end point. As before, away from extremality, $T_{\rm max}^{\rm ws}$ and $T_{\rm min}^{\rm ws}$ will be exist along the directions where $\vec{\Omega}_{1} + \vec{\Omega}_{2}$ is maximized and minimized. Here $\vec{\Omega}_{{1,2}}$ correspond to the angular velocities along the Cartans of the SO$(4)$.

\section{Conclusion \& Discussion}
\label{conclusion_discussion}
We have analysed the case of extremal chaos by studying the near horizon or the throat region of extremal and near extremal BTZ. We have done this with an aim of understanding the effect of rotation on the Lyapunov index $\lambda_L$ as seen at the conformal boundary of BTZ in \cite{Poojary:2018eszz,Jahnke:2019gxr}.
We find that in the near horizon region, the JT model captures the contribution of what we term as thermal modes towards chaos- these account for the $\lambda_L=2\pi/\beta<\lambda_L^-$ and only contribute away from extremality. These are the well studied near extremal $AdS_2$ re-parametrizations.
The modes that survive extremality have nice description in terms one of the (left-moving) $AdS_3$ PBH diffeomorphisms in the near horizon region. We are able to ascertain the contribution of these extremal modes exactly at extremality in the throat region where the above mentioned thermal modes are also present. The extremal modes seem to give rise to a $\lambda_L^-=2r^+$ at extremality, while the thermal modes contribute $\lambda_L^+=\lambda_L=0$ as expected. We also derive an effective action in the 3-dim throat region and find that the contribution from the thermal modes is important in determining the contribution from the extremal modes towards chaos. It would be interesting to obtain a JT like prescription for analysing the extremal modes' contribution away from extremality in the throat region.
\\\\
As a separate check we analyse the temperature as seen by the 2-dim worldsheet of a string probing the BTZ geometry as a function of its end point's angular velocity. We find that when the Killing horizon and the event horizon of the world-sheet metric coincides, the worldsheet temperature sees a combination of the left and the right moving temperatures. As a function of angular velocity $\omega$ the world-sheet temperature lies between $\sqrt{2}T_R<T^{\rm ws}<\sqrt{2}T_L$ with $2\pi T_L=\lambda_L^-$ \& $2\pi T_R=\lambda_L^+$. The extremes in the worldsheet temperature occurs when the string end point rotates at the speed of light. On the other hand, $T^{\rm ws}=T_H$ when the end point rotates with $\omega=r_-/r_+$ cancelling the rotation of the black hole. Note that, this probing calculation, at extremality, comes with a subtle upper bound for the angular velocity. This upper bound exists in higher dimensional AdS-Kerr geometries as well and therefore it may be connected to a generic physics at the extremal limit. This will be a particularly interesting feature to explore further, specially a potential connection with instabilities of extremal geometries (such as superradiance) will be very interesting. 
\\\\
We next proceed to analyse the near horizon region of extreme Kerr-$AdS_4$ and find that the JT action accounts for the contribution coming from the $AdS_2$ factor in the warped $AdS_3$.
We see that an analogue of the BTZ extremal modes can be seen in the warped near horizon $AdS_3$ in extremal Kerr geometry. These are the same `large' diffeomorphisms studied in the Kerr/CFT literature. The existence of a Schwarzian like behaviour can be inferred from the asymptotic symmetry algebra of $Vir\times U(1)$, where the $Vir$ corresponds to the extremal modes. Comparing the NHEK $AdS_3$ at a fixed $\theta=\theta_0$ with that of the near horizon extreme BTZ metric one can obtain a $T_L$ given by \eqref{Kerr_ext_temp}, which matches the Frolov-Thorne temperature seen by the modes outside the horizon. Here the $U(1)$ corresponds to time translations which is indeed generalized to $\tau\rightarrow f(\tau)$ when describing the thermal modes in the $AdS_2$ in the JT model. Further analysis even at extremality would require us to know a possible non-linear completion of the extremal and thermal `large' diffeomorphisms in the 4 dimensional NHEK region. In order to find the on-shell action we would have to write the family of solutions in Fefferman-Graham form asymptotically at the throat boundary. One would also have to impose relevant boundary conditions and consistent boundary terms to the action for allowing the same.  
\\\\
We also analyse the probe string world-sheet temperature for extremal Kerr-$AdS_4$, and the physics is qualitatively similar to what is already observed in rotating BTZ geometry. In fact, these feature survive in general dimensions and, on the worldsheet, one always finds the analogue of a ``left-moving" and ``right-moving" temperatures, and correspondingly Lyapunov exponents. These two limiting temperatures are essentially the maximum and the minimum temperatures that the worldsheet fluctuations observe. In the extremal limit, the worldsheet temperatures become a non-trivial function of the Frolov-Thorne temperatures. Thus, at extremality, the probe degrees of freedom display an ergodic growth of the OTOCs. It will certainly be very interesting to extract an effective description for the same, which we leave for future work. 
\\\\
Extremal black holes enjoy a better description of their microstates in terms of stringy (brane) configurations satisfying certain BPS conditions \cite{Lin:2004nb}. It would be interesting to see how such configurations individually contribute towards the chaotic process involved in scrambling of small perturbations to extremal black hole geometries. 

\section{Acknowledgements}

We would like to thank Shouvik Datta, Dileep P.~Jatkar, Lata Kh Joshi, R. Loganayagam, S. Prabhu, Junggi Yoon for various conversations related to the work here. RP would Like to thanks the hospitality of CMI and Amitabh Virmani during initial parts of this work.

\appendix
\section{Robert's Transformations}
\label{Appendix_A}
Here we list the full non-linear diffeomorphisms which map one stationary BTZ solution to another. These were first written down in the form given below by Roberts \cite{Roberts:2012aq}.
We begin with
\be
ds^2=\frac{du^2}{u^2}-u^2dy^+dy^-
\ee
which can be re-labelled to
\be
u=r\frac{8f'_+f'_--2r^{-2}f''_+f''_-}{(4f'_+f'_-)^{3/2}},\hspace{0.5cm}y^\pm=f'_\pm+\frac{4r^{-2}{f'_\pm}^2 f''_\mp}{8f'_\pm f'_\mp-2r^{-2}f''_\pm f''_\mp}
\label{roberts_diff}
\ee
\bea
\implies&& ds^2=\frac{dr^2}{r^2}-r^2dx^+dx^+\frac{1}{4}	\left(T_{++}(dx^+)^2+T_{--}(dx^-)^2\right)-\frac{1}{16r^2}T_{++}T_{--}dx^+dx^-\cr&&\cr
&&{\rm where}\,\,T_{\pm\pm}=-2\,{\rm Sch}[f^\pm(x^\pm),x^\pm]\,\,\&\,\, f_\pm\equiv f_\pm(x^\pm) .
\label{Poincare_AdS3}
\eea
where ${\rm Sch}[f(t),t]=\frac{2f'f''-3{f'''}^2}{2{f'}^2}=-\{f,t\}/2$. 
Note as $r\rightarrow \infty$, $y^\pm\rightarrow f_\pm(x^\pm)$.  
\\\\
If we were to parametrize the conformal transformations as $y\rightarrow e^{\sqrt{L_\pm}f_\pm}$ at the boundary instead, we would get:
\be
T_{\pm\pm}=\{f_\pm(x^\pm),x^\pm \}+L_\pm {f'_\pm}^2
\label{Sch_shift}
\ee
which for $f_\pm(x^\pm)=x^\pm$ implies the stationary BTZ family of metrics 
\bea
\frac{ds^2}{l^2}&=&\frac{dr^2}{r^2}-r^2dx^+dx^-+\frac{1}{4}\left(L_+dx^{+2}+L_-dx^{-2}\right)-\frac{1}{16r^2}L_+L_-dx^+dx^-,
\label{Banados_metric_1}
\eea
with $L_\pm=(r_+\mp r_-)^2=\lambda_L^{\pm 2}$. Here $r_\pm$ are inner and outer horizons in standard BTZ coordinates in \eqref{BTZ_metric_wiki}. If one were to apply infinitesimal Brown-Henneaux diffeomorphisms to the above metric the change in the $L_\pm$ would be precisely be given by linear terms in $\epsilon_\pm$ in \eqref{Sch_shift} where $f_\pm(x^\pm)=x^\pm+\epsilon_\pm(x^\pm)$. Therefore the full non-linear diffeomorphism that corresponding to the infinitesimal Brown-Henneaux or the PBH ones takes $\L_\pm \rightarrow T_{\pm\pm}$ given by \eqref{Sch_shift}. One can find the full diffeomorphisms which achieves this by inverting the coordinate map that takes us from \eqref{Poincare_AdS3} to \eqref{Banados_metric_1} and then implementing the map \eqref{roberts_diff} but with $f_\pm\rightarrow e^{\sqrt{L_\pm}f_\pm}$.  	
\\\\
We exhibit this for the simpler case of $L_+=0$ for which \eqref{Banados_metric_1} takes the form
\bea
\frac{ds^2}{l^2}&=&\frac{dr^2}{r^2}-r^2dx^+dx^-+\frac{1}{4}L_-dx^{-2},
\label{Banados_metric_Lm}
\eea
which is similar to the near horizon metric for extremal BTZ after scaling the radial coordinate.	Upon the following coordinated transformation
\be
r=\frac{u}{{f_-'(y^-)}^{3/2}},\hspace{0.5cm}x^+=y^++\frac{f_-''(y^-)}{2u^2f_-'(y^-)},\hspace{0.5cm}x^-=f_-(y^-)
\label{robets_chiral}
\ee
we get
\bea
\frac{ds^2}{l^2}&=&\frac{du^2}{u^2}-u^2dy^+dy^-+\frac{1}{4}\left(\{f_-(y^-),y^-\}+L_- {f'_-}^2\right)dy^{-2},
\label{Banados_metric_Lm_family}
\eea
The 2-dim analogue parametrizing the space of $AdS_2$ metric can be easily constructed from \eqref{roberts_diff} by taking $f_\pm(x^\pm)\rightarrow f(t)$. One can similarly obtain the Schwarzian derivative by parametrizing the space of $AdS_2$ metrics in terms of conformal transformations of $t$ at the conformal boundary of $AdS_2$. Proceeding in a similar manner as shown above a corresponding shift in the Schwarzian can also be obtained if the family of $AdS_2$ metrics are parametrized about a thermal $AdS_2$.   
\bibliographystyle{JHEP}
\bibliography{bulk_syk_soft_modes.bib}
\end{document}